\documentclass[12pt,preprint]{emulateapj} 
\usepackage{graphicx}
\usepackage{epstopdf}

\newcommand{\simlt}{\mathrel{\hbox{\rlap{\hbox{\lower4pt\hbox{$\sim$}}}\hbox{$<$}}}}
\newcommand{\simgt}{\mathrel{\hbox{\rlap{\hbox{\lower4pt\hbox{$\sim$}}}\hbox{$>$}}}}

\newcommand{\SNeIa}{SNe~Ia}
\newcommand{\SNIa}{SN~Ia}

\shorttitle{ROTSE-IIIb \SNeIa\ Census}
\shortauthors{Quimby et al.}

\begin{document}

\title{On the Rates of Type~Ia Supernovae in Dwarf and Giant Hosts with ROTSE-IIIb}

\author{
  Robert M. Quimby\altaffilmark{1},
  Fang Yuan\altaffilmark{2,3},
  Carl Akerlof\altaffilmark{\,4},
  J. Craig Wheeler\altaffilmark{5},
  \&
  Michael S. Warren\altaffilmark{6}
}

\altaffiltext{1}{
  Kavli IPMU, University of Tokyo, 
  5-1-5 Kashiwanoha, 
  Kashiwa-shi, Chiba, 277-8583, Japan
}
\altaffiltext{2}{
    Research School of Astronomy and Astrophysics, 
    The Australian National University,
    Weston Creek, ACT 2611, Australia
}
\altaffiltext{3}{
 ARC Centre of Excellence for All-sky Astrophysics (CAASTRO) 
}
\altaffiltext{4}{
  Physics Department, 
  University of Michigan, 
  Ann Arbor, MI 48109, USA
}
\altaffiltext{5}{
  Department of Astronomy, McDonald Observatory, 
  University of Texas, 
  Austin, TX 78712
}
\altaffiltext{6}{
  Theoretical Division,
  Mail Stop B227, 
  Los Alamos National Laboratory, 
  Los Alamos, NM 87545
}

\begin{abstract}

We present a sample of 23 spectroscopically confirmed Type~Ia
supernovae that were discovered in the background of galaxy clusters
targeted by ROTSE-IIIb and use up to 18 of these to determine the
local ($\overline{z} = 0.05$) volumetric rate. Since our survey is
flux limited and thus biased against fainter objects, the
pseudo-absolute magnitude distribution (pAMD) of \SNeIa\ in a given
volume is an important concern, especially the relative frequency of
high to low-luminosity \SNeIa. We find that the pAMD derived from the
volume limited Lick Observatory Supernova Search (LOSS) sample is
incompatible with the distribution of \SNeIa\ in a volume limited
($z<0.12$) sub sample of the SDSS-II. The LOSS sample requires far
more low-luminosity \SNeIa\ than the SDSS-II can accommodate.  Even
though LOSS and SDSS-II have sampled different \SNeIa\ populations,
their volumetric rates are surprisingly similar. Using the same model
pAMD adopted in the SDSS-II \SNeIa\ rate calculation and excluding two
high-luminosity \SNeIa\ from our sample, we derive a rate that is
marginally higher than previous low-redshift determinations. With our
full sample and the LOSS pAMD our rate is more than double the
canonical value. We also find that 5 of our 18 \SNeIa\ are hosted by
very low-luminosity ($M_B > -16$) galaxies, whereas only 1 out 79
nearby SDSS-II \SNeIa\ have such faint hosts.  It is possible that
previous works have under-counted either low luminosity \SNeIa,
\SNeIa\ in low luminosity hosts, or peculiar \SNeIa\ (sometimes
explicitly), and the total \SNeIa\ rate may be higher than the
canonical value.

\end{abstract}


\section{Introduction}

Calculating the rate of Type Ia supernovae (\SNeIa) and its dependence
on redshift illuminates an important contributor to the metal
enrichment history of the universe, the production rate of a specific
cross section of stellar systems, and, when linked to star formation
history, it can reveal the nature and relative fractions of the
progenitor systems. Theory and recent empirical evidence argues that
the explosions are derived from degenerate, hydrogen depleted stars,
namely white dwarfs \citep{hoyle_fowler1960,nugent2011}. The nature of
the companion star which must be present to feed mass onto these
otherwise stable constructions, however, is more controversial.

It could very well be that there are two or more pathways leading to
what we observe as \SNeIa\ including events from single degenerate
progenitors (e.g. a white dwarf accreting from a red giant star;
\citealt{whelan_iben1973}) and double degenerate progenitors (i.e. two
white dwarfs; \citealt{webbink1984}). This is a potential obstacle for
the further use of \SNeIa\ as cosmological probes if the local
population used to calibrate the peak magnitude to light curve width
relation differs systematically from the distant, cosmologically
significant population \citep[i.e. if high redshift events obey a
  different relation, cosmology studies may be
  biased;][]{domingeuz2001,mannucci2006,sullivan2006,howell2007,quimby2007a}.

The \SNeIa\ rate is one tool for resolving this issue. Locally, the
quantity of progenitor systems demanded by the \SNeIa\ rate may be
compared to the actual supply available
\citep[e.g.][]{nelemans2005,kilic2012,badenes2012}. At greater
distances, the delay time distribution (DTD), the distribution of
\SNeIa\ progenitor systems lifespans from birth to explosion, can be
compared to the distributions expected form various progenitor models
\citep{yungelson_livio2000}. The DTD can be recovered using the
\SNeIa\ rate and star-formation histories of a targeted population of
hosts \citep[][]{maoz2011a}, or by comparing \SNeIa\ over a range of
redshifts to the cosmic star formation history
\citep[e.g.][]{mannucci2006,graur2011}. The results are consistent
with a power-law distribution that favors short delays (for a review,
see \citealt{maoz2011b}).

Indeed, \SNeIa\ appear to occur more frequently in late type galaxies
with active star formation than in ellipticals
\citep[e.g.][]{mannucci2005,sullivan2006}.  Recently, the Lick
Observatory Supernova Search (LOSS) reported that the rate of
\SNeIa\ per unit luminosity of the hosts is higher for low luminosity
galaxies than luminous ones \citep{li2011b}. This so-called
``rate-size'' relation may be connected to a metallicity effect
\citep{kobayashi_nomoto2009,kistler2011}, since lower luminosity hosts
tend to have lower global metallicities
\citep[][]{tremonti2004}. White dwarfs that form in low
metallicity environments may have a mass distribution that is biased
to more massive objects as compared to solar metallicity populations
\citep[][]{umeda1999}, and this in turn may lead to a greater
supply of \SNeIa\ progenitors.

Measurement of the \SNeIa\ rate was originally performed through
searches targeting specific galaxies, and thus the rates derived were
in units of the number of \SNeIa\ found per time per galaxy
\citep{zwicky1938}. This evolved into a rate per galaxy
luminosity--SNuB or SNuK, which give the number of \SNeIa\ per
$10^{10} L_{\sun}$ (in the B or K-bands, respectively), per
century. \citet{cappellaro1999} combined targeted photographic and
visual surveys and found a \SNeIa\ rate of SNuB$=0.2\pm0.06$ averaged
over all galaxy types. LOSS finds roughly compatible \SNeIa\ SNuB
rates in specific host types if they do not include the rate-size
relation \citep{li2011b}. They also use an adopted K-band galaxy
luminosity function \citep{kochanek2001} to convert their measurements
into volumetric rates.

Methods to discover \SNeIa\ without preference for host galaxies
\citep[e.g.][]{hamuy1993,perlmutter1995b}, have been widely adopted as
the availability of wide field cameras has increased (e.g. SNLS,
ESSENCE, SDSS-II, PTF, PanSTARRS, SkyMapper). \SNeIa\ rates from such
surveys are most readily reported in volumetric units,
\SNeIa\,Mpc$^{-3}$\,yr$^{-1}$. Included in this survey category is the
SDSS-II, which was designed to bridge the gap between the local
\SNeIa\ and the high redshift events used as cosmological probes
\citep[][]{frieman2008}. Over a three year period, the SDSS-II
spectroscopically confirmed several hundred \SNeIa\ in the range $0
\simlt z \simlt 0.4$. \citet{dilday2010} use a sub sample of (normal)
\SNeIa\ discoveries to derive the $z<0.3$ rate, which is seen to
increase roughly as a power-law with redshift, $\mathcal{R} \propto
(1+z)^{2}$. \SNeIa\ rate studies have been conducted at even higher
redshifts, including work done with the Supernova Legacy Survey (SNLS;
\citealt{neill2006,perrett2012}), {\it HST} \citep{dahlen2008}, and Subaru
\citep{graur2011}. The measured \SNeIa\ rates increase to $z\sim1$,
and then level off. For a recent compilation of volumetric
\SNeIa\ rates from various surveys, see \citet{graur2011}.

In this paper, we determine the volumetric \SNeIa\ rate and
contributions to this from dwarf and giant host galaxies based on
discoveries from the ROTSE-IIIb telescope. Section \ref{sample}
discusses how our sample was selected. We discuss our unfiltered
magnitude system and how this may relate to \SNeIa\ pseudo-absolute
magnitude distributions (pAMDs) constructed from filtered observations
in \S\ref{LF}. We also discuss the pAMDs derived from the LOSS and
SDSS-II in this section. For the later, we determine an empirical pAMD
based on a volume limited sub-sample of the SDSS-II after discussing
the pAMD model that was actually used in the SDSS-II \SNeIa\ rate
calculation (and which we adopt for comparison to the SDSS-II). In
\S\ref{DE} we study our ability to detect point sources of various
brightness as a function of seeing and limiting magnitudes, and we use
this in \S\ref{survey_eff} to determine our overall survey
efficiency--the probability of discovering \SNeIa\ drawn from a given
population--as a function of distance. Next, we measure multi-band
photometry for the host galaxies of our supernovae and calculate rest
frame absolute magnitudes in \S\ref{hosts}. \SNeIa\ rates are
calculated in \S\ref{rates}, and final conclusions are offered in
\S\ref{conclusions}. Throughout this paper, we assume a flat,
$H_0=71$\,km\,s$^{-1}$\,Mpc$^{-1}$, $\Omega_m=0.27$ cosmology.

\section{Sample}\label{sample}

\begin{deluxetable*}{ccccccccccc}
\tablewidth{0pt}
\tablecaption{Background \SNIa\ Discovered with ROTSE-IIIb by Feb. 1, 2009}
\tabletypesize{\scriptsize}

\tablehead{
  \colhead{IAU Name} &
  \colhead{Disc. Date} &
  \colhead{RA} &
  \colhead{DEC} &
  \colhead{z} &
  \colhead{Disc.} &
  \colhead{$N_{\rm det}$} &
  \colhead{Abs.} &
  \colhead{Used?} &
  \colhead{Used?} &
  \colhead{Note} \\
  \colhead{} &
  \colhead{} &
  \colhead{} &
  \colhead{} &
  \colhead{} &
  \colhead{Mag} &
  \colhead{} &
  \colhead{Mag} &
  \colhead{LOSS} &
  \colhead{SDSS-II} &
  \colhead{} 
}
\startdata
2004gu & Dec 13, 2004 & 12:46:24.7 & +11:56:56 &  0.05 & 17.4 & 6  & $-$19.4 & y & y &  \\ 
2005bg & Mar 28, 2005 & 12:17:17.1 & +16:22:17 &  0.02 & 17.1 & 10 & $-$19.0 & y & y &  \\ 
2005ck & Jun 05, 2005 & 13:02:18.7 & +28:20:44 &  0.09 & 18.6 & 2  & $-$19.9 & n & n & $^{a}$  \\ 
2005cr & Jun 24, 2005 & 12:22:17.1 & +12:23:49 &  0.02 & 16.1 & 10 & $-$19.1 & y & y & $^{c}$  \\ 
2005hj & Oct 30, 2005 & 01:26:48.3 &$-$01:14:17&  0.06 & 17.6 & 17 & $-$19.7 & y & y & $^{b}$  \\ 
2005ir & Nov 03, 2005 & 01:16:43.7 & +00:47:40 &  0.08 & 18.5 & 6  & $-$19.4 & y & y & $^{b}$  \\ 
2006an & Feb 21, 2006 & 12:14:38.7 & +12:13:47 &  0.06 & 18.0 & 2  & $-$19.4 & n & n &  \\
2006cj & May 17, 2006 & 12:59:24.5 & +28:20:51 &  0.07 & 17.9 & 11 & $-$19.5 & y & y & $^{c}$  \\ 
2006ct & May 25, 2006 & 12:09:56.8 & +47:05:45 &  0.03 & 17.5 & 11 & $-$18.6 & y & y &  \\ 
2007if & Aug 16, 2007 & 01:10:51.4 & +15:27:40 &  0.07 & 19.5 & 12 & $-$20.6 & y & n & $^{d}$  \\ 
2007kh & Sep 07, 2007 & 03:15:12.1 & +43:10:13 &  0.05 & 18.9 & 10 & $-$19.3 & y & y &  \\ 
2007op & Nov 04, 2007 & 01:53:12.4 & +33:44:34 &  0.09 & 18.5 & 6  & $-$19.1 & y & y &  \\ 
2007qc & Oct 27, 2007 & 11:57:04.7 & +53:29:55 &  0.04 & 16.8 & 11 & $-$19.1 & y & y &  \\ 
2007sp & Nov 14, 2007 & 12:04:42.3 & +49:11:09 &  0.02 & 16.8 & 7  & $-$18.2 & y & y & $^{e}$ \\ 
2007sw & Dec 29, 2007 & 12:13:36.9 & +46:29:36 &  0.03 & 16.0 & 2  & $-$18.8 & n & n &  \\ 
2008E  & Jan 04, 2008 & 11:25:37.0 & +52:08:26 &  0.03 & 18.2 & 18 & $-$18.7 & y & y &  \\ 
2008ab & Jan 30, 2008 & 11:34:45.9 & +53:57:51 &  0.07 & 18.2 & 8  & $-$19.7 & y & n &  \\ 
2008ac & Jan 30, 2008 & 11:53:45.2 & +48:25:22 &  0.05 & 17.6 & 15 & $-$19.2 & y & y &  \\ 
2008ad & Jan 30, 2008 & 12:49:37.2 & +28:19:47 &  0.05 & 18.0 & 2  & $-$19.1 & n & n &  \\ 
2008ar & Feb 27, 2008 & 12:24:37.9 & +10:50:17 &  0.03 & 16.9 & 17 & $-$18.8 & y & y &  \\ 
2008bg & Mar 12, 2008 & 12:51:11.9 & +26:17:40 &  0.06 & 18.7 & 9  & $-$19.3 & y & y & $^{c}$  \\ 
2008by & Apr 19, 2008 & 12:05:21.0 & +40:56:46 &  0.05 & 17.2 & 2  & $-$19.8 & n & n &  \\ 
2008bz & Apr 22, 2008 & 12:38:57.7 & +11:07:46 &  0.06 & 17.7 & 6  & $-$19.2 & y & y &  \\ 
\enddata
\tablecomments{
$^{a}$Also reported by LOSS \citep{pugh2005}.
$^{b}$Also detected by SDSS-II and used in their volumetric rate calculations \citep{dilday2008,dilday2010}.
$^{c}$Spectroscopically confirmed by the CfA \citep{modjaz2005,colesanti2006,yuan2008b}.
$^{d}$Also reported by the Nearby SN Factory \citep{yuan2007,scalzo2010}.
$^{e}$Maximum light not well constrained.
Magnitudes given are in the ROTSE-III bandpass calibrated against the
USNO-B1.0 R2 mags as described in the text. The $N_{\rm det}$ column
gives the number of times each object was selected by the automated
search pipeline. The ninth and tenth columns indicate whether the
object was included in the rate calculations using the LOSS pAMD or
the SDSS-II model pAMD, respectively.  }
\label{table:disco}
\end{deluxetable*}

Our sample is drawn from the background population of supernovae
discovered by ROTSE-IIIb in the course of the Texas Supernova Search
(TSS; \citealt{quimby_phd}) and the ROTSE Supernova Verification
Project (RSVP; \citealt{yuan2007b}). Details of the full ROTSE-IIIb
supernova sample will be presented in a forthcoming paper. Here we
summarize the key characteristics of the sample selection process.

The TSS began survey operations in November of 2004 and was succeeded
by RSVP in early 2008. Our survey instrument, ROTSE-IIIb, has a
0.45\,m aperture and a $1.85 \times 1.85$ degree field of view. The
search continues to operate and has been expanded to the other
ROTSE-III telescopes in Australia, Namibia, and Turkey, but for this
work we consider only discoveries made by ROTSE-IIIb in Texas prior to
February 2009. 

\begin{figure}
\includegraphics[width=\linewidth]{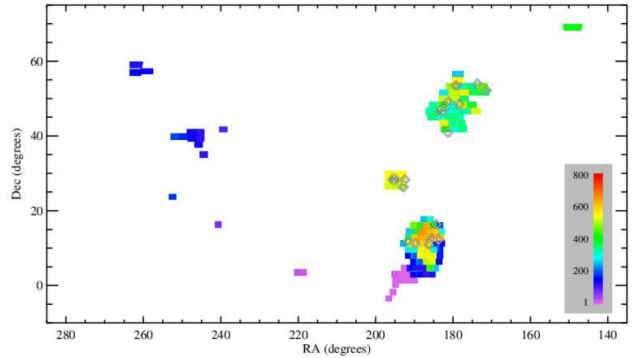}
\caption{ Winter-Spring search fields for the ROTSE-IIIb surveys. The
  color scale indicates the total number of nights each field was
  surveyed (i.e. reference epochs are not included). The diamonds mark
  the locations of background \SNeIa. }
\label{fig:winter_spring}
\end{figure}

\begin{figure}
\includegraphics[width=\linewidth]{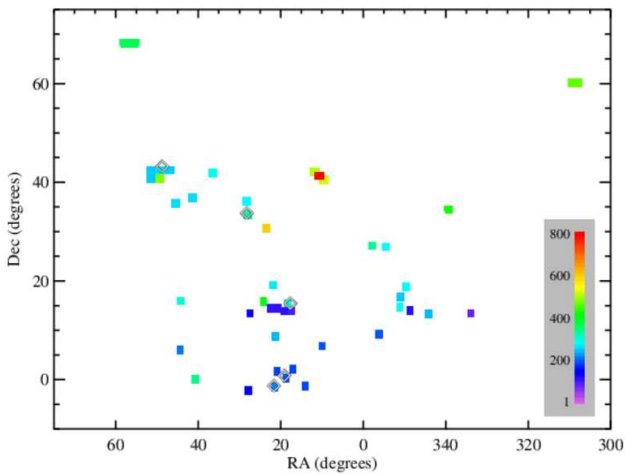}
\caption{ Similar to Figure~\ref{fig:winter_spring} but for the
  Summer-Fall fields. }
\label{fig:summer_fall}
\end{figure}

We typically targeted galaxy clusters such as Virgo, Ursa Major, Coma,
and Abell Clusters for our supernova search, but in a few exceptional
cases we chose to target specific galaxies (e.g. M31). In the present
work, we remove this potential source of bias by selecting only the
subset of \SNeIa\ found in the background of each field ($z > z_{f} +
4000$\,km\,s$^{-1}$, where $z_f$ is the redshift of the object
targeted in a given field).  During the search period, we detected 76
supernovae in total, and of these 46 were classified as \SNeIa. After
removing foreground and cluster supernovae, we are left with the 23
\SNeIa\ background events listed in Table~\ref{table:disco}.  Figures
\ref{fig:winter_spring} and \ref{fig:summer_fall} show our sky
coverage with the locations of the discoveries marked. Fields with the
highest concentrations of nearby galaxies were typically observed
every night as weather and season allowed. Over the first few years we
observed the less rich fields on alternating nights. Additional time
was allocated for the survey in the RSVP era, and fields were added at
that time. 

For both TSS and RSVP, we employed image subtraction techniques to
reveal time variable events. The TSS employed a modified version of
the \citet{perlmutter1999} search code, and the RSVP used the process
described in \citet{ya2008}.  For our survey, we typically observe
each field multiple times each night. From these we create 3 co-added
frames representing the average of the first half of the nights data
(NEW1), the average of the second half (NEW2), and the average for the
whole night (NEW). We use filtering to remove particle events and
other artifacts during the co-addition. We then subtract the properly
convolved reference template from each of these to generate subtracted
frames (SUB1, SUB2, and SUB). To remove contamination from solar
system bodies and remaining artifacts, we require a candidate to be
detected ($\simgt2.5\sigma$) at consistent celestial coordinates on
each of SUB1 and SUB2. The candidate is finally required to be
detected at the $5\sigma$ confidence level on the combined
subtraction, SUB. The typical FWHM for the survey was 3.2 pixels
(about 10'') and the $3\sigma$ limiting magnitude for the nightly
coadds was typically around 18.3 mag.

We further vet potential SN candidates against catalogs of known
time-variable phenomena\footnote{For example:
  \url{http://scully.cfa.harvard.edu/cgi-bin/checkmp.cgi},
  \url{http://simbad.u-strasbg.fr/simbad/sim-fid},
  \url{http://nedwww.ipac.caltech.edu/}}. Most of our fields were
observed by the SDSS as well, so we were able to remove contamination
from un-cataloged variable stars by checking for matching ``host''
stars in the SDSS data, which is complete to $\sim4$ magnitudes below
our survey depth. Remaining sources were vetted by human scanners. All
sources judged as possible transients were spectroscopically
classified by us or others in the community.  This is a key
distinguishing feature of our survey; other wide-field transient
searches generate more candidate supernovae than can be classified
with the spectroscopic resources available, so a filtering process
such as light curve fitting is typically employed to select only a
subset for final spectroscopic confirmation. 

Table~\ref{table:disco} also notes the number of times that each
discovery was recovered by the automated search pipeline ($N_{\rm
  det}$). A final analysis of the data adds low significance
(i.e. $<5\sigma$) detections that are not passed by the automated
selection cuts, and we determine the peak magnitudes for each
supernova in our sample by fitting light curve templates to these
final data sets. We use R-band light curve templates for \SNeIa\ from
LOSS, which were constructed by interpolating between the faint/fast
SN~1999by and the bright/slow SN~1991T \citep{li2011a}. See \S\ref{LF}
for a discussion of our unfiltered spectral response and a comparison
to the R-band. We use {\tt mpfit.pro} in IDL to determine the best
fitting template shape, and subtract off the Galactic extinction in
the R-band \citep{sfd1998}. We employ the distance modulus (based on
independent distance indicators when available through
NED\footnote{\url{http://ned.ipac.caltech.edu/}} or otherwise based on
the redshift in a flat, $H_0=71$\,km\,s$^{-1}$\,Mpc$^{-1}$,
$\Omega_m=0.27$ cosmology) to determine the peak absolute magnitudes
shown in table~\ref{table:disco}. SN~2004gu was first observed near
peak, which adds some uncertainty to the date and magnitude of maximum
light; however, spectroscopic phase information is consistent with the
derived date of maximum light, and thus we expect the peak magnitude
estimate to be accurate. According to its spectroscopic age, another
event, SN2007sp, was first detected about two months after maximum
light, so the light curve fit likely underestimates the peak.

\subsection{Sample Notes}\label{sample_notes}

Although our sample is small, it contains several \SNeIa\ worth of
special note.
\begin{itemize}
\item {\bf SN~2004gu} is a significant outlier on the Hubble diagram,
  and other authors have grouped it with peculiar events like
  SN~2006gz and SN~1999ac, which may or may not be linked to
  super-Chandrashekar explosions
  \citep[][]{contreras2010,silverman2011}.

\item {\bf SN~2006ct} showed the characteristic features of a Type~Ia
  supernova, but the blue-shift of the line minima near maximum light
  was only $\sim6000$\,km\,s$^{-1}$ \citep{quimby2006}. The spectra
  show some similarities to the highly peculiar SN~2002es
  \citep[][]{ganeshalingam2012}, which was also hosted by an early
  type galaxy, but the photometric decline of SN~2006ct
  \citep{quimby_phd} is slower and more reminiscent of SN~2002cx
  \citep{li2003a} and SN~2005hk \citep{phillips2007}.

\item {\bf SN~2007if} is the most luminous Type~Ia supernova known,
  and it may be the result of a super-Chandrashekar mass explosion
  \citep{scalzo2010,yuan2010}.

\item {\bf SN~2007qc} was hosted by an extremely low luminosity dwarf
  galaxy (see \S\ref{hosts}). At $M_B \sim -11$, the host is among the
  faintest detected for any \SNeIa.

\item {\bf SN~2008ar} showed spectroscopic evidence for high velocity
  material and unusually strong absorption in the \ion{Ca}{2} IR
  triplet \citep{yuan2008a}, which is reminiscent of SN~2007le
  \citep{simon2009}.
\end{itemize}
 
In addition, at least 7 of our \SNeIa\ (2004gu, 2005ck, 2005hj,
2006cj, 2007if, 2008ab, 2008by and possibly 2005bg) show either
unusually high luminosities or are spectroscopically classified as
SN~1991T/1999aa-like. This appears to be a high fraction even for a
magnitude limited survey when compared to the LOSS census of
\SNeIa\ in targeted galaxies \citep{li2011a}.

\section{\SNIa\ pseudo-Absolute Magnitude Distributions}\label{LF}

Our survey is flux limited, and as a result we have a bias against
selecting low luminosity \SNeIa. This bias is exacerbated by our
removal of discoveries made at or below the redshift of the targeted
galaxy clusters. To determine the event rate for the
\SNeIa\ population as a whole, we must therefore correct the number of
events in our observed sample for the fraction we are likely to have
missed due to, among other concerns, their lower luminosities (see
\S\ref{survey_eff}). This requires knowledge of the intrinsic
luminosity function for all \SNeIa\ in our search volume combined
with the host galaxy absorption distribution. 

In this section, we consider the pseudo-absolute magnitude
distribution (pAMD) for \SNeIa, which is the distribution one would
obtain by correcting the peak observed magnitudes of a complete sample
of \SNeIa\ for distance and Galactic extinction but not for host
galaxy extinction. This distribution can directly be used to test the
selection efficiency of a flux limited survey (see
\S\ref{survey_eff}). First, we consider in
\S\ref{filtered_v_unfiltered} how the available \SNeIa\ luminosity
functions derived (mostly) through filtered observations may compare
to our unfiltered survey data. We then discuss the pAMD compiled by
LOSS from their targeted, but volume limited sample \citep{li2011a} in
\S\ref{lossLF}, and we derive an empirical pAMD from a volume limited
sub sample of the SDSS-II survey in \S\ref{sdssLF}. We find that the
LOSS and SDSS-II pAMDs are not consistent, which is discussed in
\S\ref{loss_v_sdss}.

\subsection{Comparison of R-band and ROTSE-IIIb Unfiltered Magnitudes}\label{filtered_v_unfiltered}

We calibrate our instrumental magnitudes against the USNO-B1.0 R2
system since some of our search fields are outside of the SDSS
footprint. For fields covered by both the SDSS and USNO-B1.0, we
compared the USNO-B1.0 photometry to SDSS photometry
converted\footnote{See the Lupton (2005) equations at
  \url{http://www.sdss.org/dr7/algorithms/sdssUBVRITransform.html}} to
the R-band using $R = r - 0.2936(r - i) - 0.1439$, and we find the two
systems agree to about 0.01\,mag on average and have a field to field
dispersion of about 0.14\,mag. Thus our unfiltered ROTSE-IIIb
magnitudes should be, on average, similar to the R-band, but objects
with spectral energy distributions that differ significantly from the
field stars employed in the calibration may be offset from their true
R-band values.

\begin{figure}
\includegraphics[width=\linewidth]{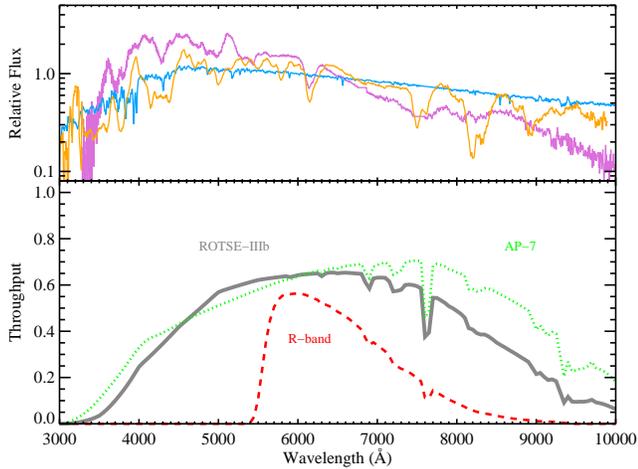}
\caption{ Determination of the magnitude offset between filtered and
  unfiltered measurements. The top panel shows the spectral energy
  distributions of a typical field standard (blue) compared to a high
  luminosity \SNIa\ (purple) and a low-luminosity
  \SNIa\ (orange). These can be convolved with the ROTSE-IIIb (grey)
  and R-band (dashed red) response curves in the lower panel to
  determine the magnitude offset. The dotted green curve is the
  typical response for an Apogee7 camera, which is used by KAIT for
  the LOSS. See text for details. }
\label{fig:bandpass}
\end{figure}

We estimate the potential offset between our unfiltered magnitudes and
the true R-band system for \SNeIa\ as follows. We adopt the spectral
energy distribution of a G5V star \citep{pickles1998} as
representative of a typical field standard used in calibrating the
photometric zeropoint of our unfiltered imaging data. We scale this
template to zero magnitude in the R-band, and then measure the
synthetic flux using the approximate transmission function of
ROTSE-IIIb \citep{quimby2007b}, which is plotted in
Figure~\ref{fig:bandpass}. We next scale a
template\footnote{\url{http://supernova.lbl.gov/~nugent/nugent\_templates.html}}
spectrum for a Branch-normal \SNIa\ at maximum light to zero magnitude
in the R-band and measure its flux in the ROTSE-IIIb bandpass. The
ratio of this and the scaled field star flux (also measured in the
ROTSE-IIIb bandpass) gives the expected departure of the ROTSE-IIIb
magnitudes from the true R-band system. We measure this offset in the
range $0.02 < z < 0.09$ by redshifting the \SNIa\ template, and find
that our unfiltered magnitudes may be 0.05 to 0.10\,mag brighter than
the true R-band system. We repeat this process for a high luminosity
\SNIa\ (SN~1999aa; \citealt{garavini2004}), and a low luminosity event
(SN~1991bg; \citealt{filippenko1992}). We find that a \SNIa\ with the
same R-band brightness as a field standard will generate a magnitude
in the unfiltered system (calibrated against the R-band) that is
within $\sim0.1$\,mag of its true R-band magnitude.

While this shows that our observed magnitudes should be similar to the
true R-band system, a more pertinent concern is how the derived
absolute magnitudes will compare. In particular, we have calculated
absolute magnitudes from our unfiltered data by simply subtracting off
the distance moduli and Galactic (R-band) extinction terms; we have
not included a ``k-correction'' term to account for the changing
rest frame bandpass of our unfiltered system over the modest redshift
distribution of our sample. Below, we compare the pseudo absolute
magnitudes derived from our data to rest frame (k-corrected) R-band
magnitudes calculated for the same sources based on filtered
photometry to check for an offset.

The CfA3 sample includes MLCS2k2 \citep{jha2007} fits for four of the
supernovae in our sample \citep{hicken2009}, including two that were
also in the SDSS-II sample \citep{kessler2009}. We calculate the
pseudo-absolute magnitudes from these fits using the $\Delta$
parameters and adding back in the host extinction term. For this
comparison, we calibrate the ROTSE-IIIb magnitudes against the SDSS
field stars (converted to the R-band as described above) to remove the
USNO-B1.0 vs. SDSS-II offset, which we find to be biased in this small
overlapping sample.

Our pseudo absolute magnitude estimate for SN~2005hj is about 0.1\,mag
brighter than the fits from the CfA3 and the SDSS-II. For SN~2005ir,
the SDSS-II and CfA3 differ from each other by about 0.1\,mag, which
is comparable to the combined measurement error. Our absolute
magnitude for SN~2005ir is about 0.02\,mag brighter than the SDSS-II
value and 0.09\,mag fainter than the CfA3 value. Our measurements of
SN~2006an and SN~2006cj are 0.03\,mag brighter and 0.16\,mag fainter
than the CfA3 peak magnitudes, respectively.

Additionally, the Nearby Supernova Factory has reported a peak V-band
magnitude and $\rm{V}-\rm{R}$ color for SN~2007if that implies a peak
absolute R-band maximum of about $-20.4$\,mag. Calibrating against the
SDSS r-band AB magnitudes, \citet{yuan2010} found the ROTSE-III
observations peaked at $M_r=-20.4$. Field stars will typically
``brighten'' by about 0.22\,mag when converting from the SDSS r-band
system to Cousins R-band (Vega) magnitudes, which shifts the target
photometry by the same amount. Our estimate for the peak of SN~2007if
is approximately 0.15\,mag brighter than the Nearby Supernova Factory
value.

To, summarize, of the 7 pseudo-absolute magnitude measurements
available in the literature based on (k-corrected) filtered
photometry, our ROTSE-IIIb estimates are $\sim0.1$\,mag brighter in
three cases, $\sim0.1$\,mag fainter in 2 cases, and roughly the same
in the remaining 2 cases.  On average, our values are 0.02\,mag
brighter. We conclude that there is no significant systematic offset
between our unfiltered, pseudo-absolute magnitude estimates and the
(mostly) R-band system that defines the luminosity functions
considered below (\S\ref{lossLF} and \ref{sdssLF}), and these should
agree to about $\sim0.1$\,mag for individual events. We address the
impact the possible systematic error in our magnitude system with
respect to the LOSS and SDSS-II pseudo-absolute magnitude
distributions will have on our rates in \S\ref{conclusions}.

\subsection{The LOSS pseudo-Absolute Magnitude Distribution for \SNeIa}\label{lossLF}

The Lick Observatory Supernova Search (LOSS) has recently published a
distribution of pseudo-absolute magnitudes for \SNeIa\ from their
volume limited search \citep{li2011a}. Like ROTSE-IIIb, the LOSS
survey engine (the Katzman Automatic Imaging Telescope, or KAIT) runs
with an unfiltered CCD imager, but follow-up observations are often
obtained through standard broadband filters. The \SNeIa\ luminosity
function reported by \citet{li2011a} is primarily constructed from
R-band data, which the authors describe as the best match to their
unfiltered survey data, and also from unfiltered observations as
well. In the lower panel of Figure~\ref{fig:bandpass} we show the
response curve for the Cousins R-band compared to the typical
unfiltered response of an Apogee7 camera, which is used by LOSS.

It is important to note that the LOSS \SNeIa\ luminosity function was
drawn from a sample that was highly biased with respect to the host
galaxy properties. In particular, the LOSS targeted mainly high
luminosity galaxies. It has been demonstrated that certain
\SNeIa\ sub-types prefer certain host galaxies
\citep[][]{sullivan2006}, so the luminosity distribution of the LOSS
sample may be biased with respect to the larger population from which
our sample is drawn.

An obvious difference between the LOSS and ROTSE-IIIb \SNIa\ samples
is the presence of high luminosity ($M_R < -19.6$) events in the
latter. Although we could follow previous works and simply discard our
most luminous \SNeIa, we instead chose to calculate an inclusive
rate. We must, therefore, augment the LOSS luminosity function to
account for such events or else our effective survey volume will be
under estimated. Lacking detailed demographics for this population, we
simply assume that the population of high luminosity
\SNeIa\ represents only about 1\% of \SNeIa\ in a given volume; if it
were much more these would be detected more frequently, and if it were
much less we could not expect to find any in our small sample (see
\S\ref{survey_eff}). We assume that SN~2007if was a particularly
luminous example of this population (as noted above, it is the most
luminous \SNIa\ known) and choose a half Gaussian with a peak at
$M_R=-19.6$ and $\sigma = 0.4$\,mag to stand in for the unknown
luminosity distribution. We will use this augmented version of the
LOSS pAMD to calculate the \SNeIa\ rate in \S\ref{rates}.

\subsection{The SDSS-II pseudo-Absolute Magnitude Distribution for \SNeIa}\label{sdssLF}

For our rate calculations in \S\ref{rates}, we will also use the same
bimodal-Gaussian luminosity function attenuated by a host galaxy
absorption distribution that was assumed by \citet{dilday2010} in
deriving the SDSS-II rate. The model is a revised version\footnote{
  Dilday, priv. comm.} of the distribution assumed by
\citet{dilday2008} that takes into account the findings of
\citet{kessler2009}. The difference makes for only a negligible change
in our rate measurment. The intrinsic \SNeIa\ luminosity function is
defined in terms of the MLCS2k2 light-curve shape/luminosity
parameter, $\Delta$. The distribution of $\Delta$ is defined as a
Gaussian with $\sigma=0.19$ for $\Delta < -0.2$ and $\sigma=0.40$ for
$\Delta > -0.2$, and it is truncated to lie within the range ($-0.4 <
\Delta < 1.8$). The final, pseudo-absolute magnitude distribution is
achieved by adding host galaxy absorption drawn from a distribution
with $P(A_R) \propto e^{-A_R/0.28}$. With our choice of $H_0$, peak
pseudo-absolute magnitudes are then defined as \citep[cf.][]{jha2007}:

\begin{equation}
M_R = -19.313 + 0.579 \Delta + 0.254 \Delta^2 + A_R
\end{equation}

When applying the SDSS-II pAMD model to our ROTSE-IIIb discoveries, it
is important to note that the SDSS-II rate is based on a sub-sample of
\SNeIa, and to be consistent with the SDSS-II pAMD model we must also
remove objects from our sample. In particular, the SDSS-II includes
only the fraction of the population that is well fit by MLCS2k2
($\mathcal{P}_{\rm fit} > 0.001$) with a light curve width parameter
$\Delta > -0.4$. Such a cut excludes high luminosity \SNeIa, and in
particular, SN~1999aa-like events \citep{kessler2009}. We note that
this is a common choice. The Supernova Legacy Survey (SNLS), for
example, has presented a volumetric rate for \SNeIa\ at $\overline{z}
\sim 0.45$, but the spectroscopically confirmed Type Ia SN~2003fg
(SNLS-03D3bb) is not included \citep{neill2006,howell2006}. Indeed,
the selection cuts that are applied in selecting candidates for
spectroscopic follow-up may also pose a bias against high luminosity
\SNeIa. We therefore remove events with peak luminosities brighter
that $M_R < -19.7$ when using the model pseudo-absolute magnitude
distribution from the SDSS-II.

We next look at the actual distribution of pseudo-absolute magnitudes
from a volume limited sub-sample of the first year SDSS-II
\SNeIa\ discoveries \citep{frieman2008,sako2008}, which can be
compared to the assumed model as well as the LOSS pAMD. According to
\citet{dilday2008}, the SDSS-II could have detected even a
sub-luminous (SN~1991bg-like) \SNIa\ out to a redshift of $z\sim0.12$,
so we assume the survey to be volume limited out to this limit. There
were 29 spectroscopically confirmed (or probable) \SNeIa\ from the
first year of the SDSS-II search within this redshift limit. In
addition, \citet{dilday2008} lists one probable \SNIa\ (SDSS-II
SN09266; based on photometric constraints) with a host redshift of
$z=0.0361$. This source is not included in the sample of
\citet{dilday2010} due to an additional cut on the photometric
screening. A second photometrically probable source, SDSS-II SN09739,
with a spectroscopic redshift for its host and a third,
photometrically probable \SNIa\ (SDSS-II SN11092) are also given for a
total of 32 \SNeIa\ in the volume limited sample.

We now derive an empirical \SNeIa\ pAMD from this volume limited sub
sample of the first-year SDSS-II. We adopt the temporal selection cuts
of \citet{dilday2008} and \citet{kessler2009} to remove events with
poorly constrained maxima, but unlike these samples, we include
peculiar events. For the bulk of the sample, we derive the peak
absolute R-band magnitudes from the MLCS2k2 \citep{jha2007} fit
parameters given in \citet{kessler2009}. We add the host extinction
estimate back in to retain the pseudo-absolute distribution required
for rate studies. Two peculiar \SNeIa, SN~2005hk and SN~2005gj, were
cut from the \citet{kessler2009} sample; it is still a matter of
debate if these events truly belong to the Type Ia class
\citep[e.g.][]{valenti2009,foley2010,maund2010,aldering2006,trundle2008}. We
include them here for completeness. For SN~2005hk, we take the LOSS
measurement for the R-band absolute magnitude \citep{li2011a}, and for
SN~2005gj we take the peak absolute r-band magnitude from
\citet{prieto2007} and convert this to R-band. We similarly use the
SDSS-II photometric measurements of \citet{holtzman2008} to estimate
peak absolute R-band magnitudes for two spectroscopically probable
\SNeIa: SN~2005je and SDSS-II SN06968. Finally, we choose to include
an estimated peak magnitude for SDSS-II SN09266. There is no published
light curve or peak magnitude for this event, so we assign it an {\it
  ad hoc} pseudo-absolute magnitude of $M_R=-16$ mag based on the
heavy extinction reported ($A_V\sim4$; \citealt{dilday2008}).

We thus have peak magnitudes for 22 of the first year SDSS-II \SNeIa;
eight of the remaining events have poorly constrained maxima. The
portions of the light curves sampled by the SDSS-II, however, imply a
similar distribution of peak magnitudes to the well constrained
sample. Photometry is not available for the other two photometrically
selected events (SDSS-II SN09739 and SN11092), so we cannot include
these. The peak R-band pseudo-absolute magnitudes (corrected to
$H_0=71$\,km\,s$^{-1}$\,Mpc$^{-1}$) for the volume limited SDSS-II sample are
listed in table~\ref{table:sdss}.

\begin{deluxetable}{rlcc}
\tablewidth{0pt}
\tablecaption{SDSS-II First Year Volume Limited Sample}
\tabletypesize{\scriptsize}
\tablehead{
  \colhead{SDSS ID} &
  \colhead{IAU Name} &
  \colhead{$M_R$} &
  \colhead{Note}
}
\startdata
  722 &  2005ed  & \nodata&     \\
  739 &  2005ef  & \nodata&     \\
  774 &  2005ei  & \nodata&     \\
 1241 &  2005ff  & -18.97 &     \\
 1371 &  2005fh  & -19.39 &     \\
 2102 &  2005fn  & \nodata&     \\
 2561 &  2005fv  & -19.03 &     \\
 3256 &  2005hn  & -19.11 &     \\
 3592 &  2005gb  & -19.30 &     \\
 3901 &  2005ho  & -19.33 &     \\
 4524 &  2005gj  & -20.47 & $^{a}$    \\
 5395 &  2005hr  & -19.36 &     \\
 5549 &  2005hx  & -19.23 &     \\
 5944 &  2005hc  & -19.41 &     \\
 6057 &  2005if  & -18.95 &     \\
 6295 &  2005js  & -17.72 &     \\
 6558 &  2005hj  & -19.36 &     \\
 6773 &  2005iu  & -19.28 &     \\
 6962 &  2005je  & -19.28 & $^{b}$    \\
 6968 &  \nodata & -19.06 & $^{b}$    \\
 7147 &  2005jh  & -18.96 &     \\
 7876 &  2005ir  & -19.29 &     \\
 8151 &  2005hk  & -18.36 & $^{c}$    \\
 8719 &  2005kp  & -19.30 &     \\
 9266 &  \nodata & -16.00 & $^{d}$    \\
 9739 &  \nodata & \nodata&     \\ 
10028 &  2005kt  & -19.06 &     \\ 
10096 &  2005lj  & \nodata&     \\
10434 &  2005lk  & \nodata&     \\
10805 &  \nodata & \nodata&     \\
11067 &  2005ml  & \nodata&     \\
11092 &  \nodata & \nodata&     \\ 
\enddata
\tablecomments{
$^{a}$Peak magnitude converted from \citet{prieto2007}.
$^{b}$Peak magnitude estimated from \citet{holtzman2008} light curve.
$^{c}$Peak magnitude from \citet{li2011a}.
$^{d}$Peak magnitude estimate based on description in \citet{dilday2008}.
Remaining magnitude estimates from \citet{kessler2009}. Missing $M_R$ values result from events discovered at the beginning or end of the search period for which maximum light is not well constrained, or because photometry is not available (9739 and 11092). Also note that the values in the $M_R$ column are pseudo-absolute magnitudes that have not been corrected for host galaxy extinction.
}
\label{table:sdss}
\end{deluxetable}

\subsection{Comparison of the LOSS and SDSS-II pseudo-Absolute Magnitude Distributions}\label{loss_v_sdss}

\begin{figure}
\includegraphics[width=\linewidth]{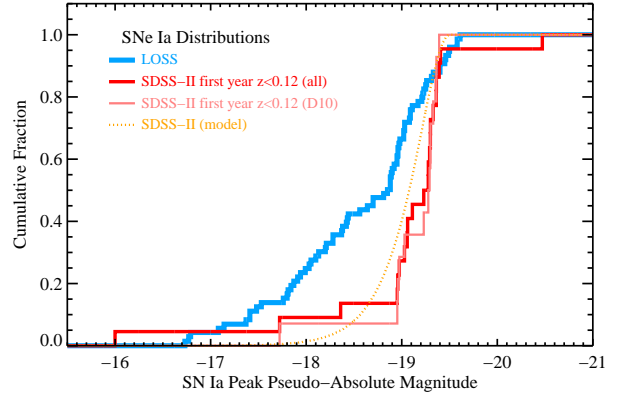}
\caption{ Absolute \SNeIa\ magnitude distributions from the LOSS
  (blue) and SDSS-II (red) volume limited samples. The pink line is
  the subset from the volume limited SDSS-II sample that was used in
  the rate calculation of \citet{dilday2010}. The dotted orange line
  is the model pAMD assumed in the SDSS-II rate calculation; we use
  this model to determine a \SNeIa\ rate that can be directly compared
  to the SDSS-II value. For the pseudo-absolute magnitudes, the
  observed magnitudes are corrected for Galactic extinction and the
  distance moduli are removed, but the host extinction is left
  uncorrected. }
\label{fig:SNeIa_LF}
\end{figure}

A comparison of the LOSS \SNeIa\ luminosity function and the SDSS-II
first year volume limited sample shows a striking result: the samples
do not agree (Fig.~\ref{fig:SNeIa_LF}). About 66\% of the LOSS
\SNeIa\ are fainter than $-19$ mag, but only about 27\% of the events
in the SDSS-II sample are this faint. We discuss the implications of
this disagreement in \S\ref{conclusions}. Only SN~2005gj, which
appears to draw its power through interaction with a hydrogen shell
and could technically be classified as a Type\,IIn, approaches the
luminosity of our higher luminosity events, such as SN~2007if
\citep{scalzo2010,yuan2010}.

Also shown in Figure~\ref{fig:SNeIa_LF} is the R-band pAMD model that
was assumed in calculating the SDSS-II \SNeIa\ rates
\citep{dilday2010}, which also appears to differ from the selected
SDSS-II population. From a KS test, we find that there is only about a
4\% chance that the volume limited SDSS-II \SNeIa\ sample is
consistent with the attenuated, bimodal-Gaussian model of
\citet{dilday2010}. The reddening distribution assumed in the model
predicts that it is extremely improbable to have highly extincted
events such as SDSS-II SN09266 (again, this was not spectroscopically
confirmed and not included in \citealt{dilday2010}). We repeated the
KS test using only the members of the volume limited sample that were
also used by \citet{dilday2010} in their rate calculation and find the
probability that these events were drawn from the assumed model is
only about 3\%. Except for a few low-luminosity events, the observed
SDSS-II sample is systematically biased to higher luminosities than
the model predicts. We discuss the SDSS-II model as applied to the
ROTSE-IIIb sample in \S\ref{rates}.

\section{Detection Efficiency}\label{DE}

In this section we determine the detectability of sources as a
function of observed magnitude relative to the limiting magnitude of a
given image. This curve will be used later to determine the
probability of detecting simulated sources with various observed
magnitudes, which is needed to calculate our overall survey
efficiency.

To determine the probability that a source of a given brightness will
be detected, we perform a Monte Carlo simulation. We add simulated
point sources to our co-added images and determine the fractions
recovered as a function of magnitude. This is done relative to the
limiting magnitude of the images and coadds recorded in the logs,
which is calculated from the 90th percentile magnitude of all objects
extracted by SExtractor \citep{bertin_arnouts1996}. Our reference
images are typically constructed from dozens of the best individual
frames from each field, so they contribute negligibly to the noise on
the subtracted frames. We therefore choose to calculate the
detectability of sources directly from blank sky regions of the
co-added images without performing the computationally expensive
process of image subtraction. A consequence of this is that we do not
account for the effects of contaminating host galaxy light. Other
surveys that employ image subtraction have previously tested the
effects of host light on detectability and found very little change
from bright hosts to blank sky \citep{neill2006}. Nonetheless, our
procedure may overestimate our detection probability for supernovae
hosted by bright galaxies (i.e. the rate may be {\it
  under}estimated). Our results do, however, have direct applications
for objects with faint hosts or otherwise dark backgrounds.

On a given image, the sensitivity to objects of a given magnitude will
vary with location in the field due to variations in the instrumental
PSF as well as structure in the atmospheric transmissivity
(e.g. passing clouds), which can be resolved by our wide field. To
account for these effects, we locate a set of isolated field standards
on each image to use in calculating the local PSF and zeropoint. We
divide each image into a grid with as many cells as can be made while
maintaining around 100 or more field standards in each (typically
$3\times3$ cells), and then calculate the average zeropoint and PSF
for each of these cells. The PSF is found using the DAOPHOT routines
in IDL, and as noted above (\S\ref{filtered_v_unfiltered}), we use the
USNO-B1.0 R2 system for our zeropoints.

To locate appropriate blank sky regions at which we may add our
simulated point sources, we first run SExtractor on each unadulterated
image with a low detection threshold to identify all objects and
artifacts in the field. We then convolve the object mask image from
SExtractor with the average PSF and select from the remaining,
unmasked pixels when placing simulated sources. We find that even the
small perturbations induced by adding a very faint ($m > m_{\rm lim} +
2$) test source can sometimes elevate a faint ($m \sim m_{\rm lim}$)
source previously ignored by SExtractor into a spurious detection. It
is therefore important to avoid locating test sources near such areas,
which is why we use a lower extraction threshold when creating the
object mask.

We use the local PSF and zeropoint determined above to determine the
shape of the test source and scale it to a randomly chosen
magnitude. Using the {\tt dao\_value} routine, we create a thumbnail
image of the source. We then add this source to the image one pixel at
a time by drawing from a Poissonian distribution with the expectation
values set by the thumbnail pixel values and accounting for the
detector gain. We finally run SExtractor with the same parameter file
used in the search. In order to simulate a statistically meaningful
number of test objects without biasing the extraction process, we
place only 20 test sources in each cell and repeat the simulation
multiple times for each image. The minimum pitch for the test objects
is set to be at least four times the average FWHM, so they do not
interfere with each other. For the survey, candidates must be detected
at a signal to noise greater than a set limit and they must have a
FWHM between half and double the image average to be considered. Both
of these quantities are effected by the noise in the image and Poisson
statistics. We therefore define our detection efficiency from the
number of simulated point sources that are detected by SExtractor with
parameters that pass these selection cuts relative to the total number
simulated. We performed this test on over 7000 co-added images made
from 2 or more frames. For comparison, our survey includes about 55000
master subtractions.

\begin{figure}
\includegraphics[width=\linewidth]{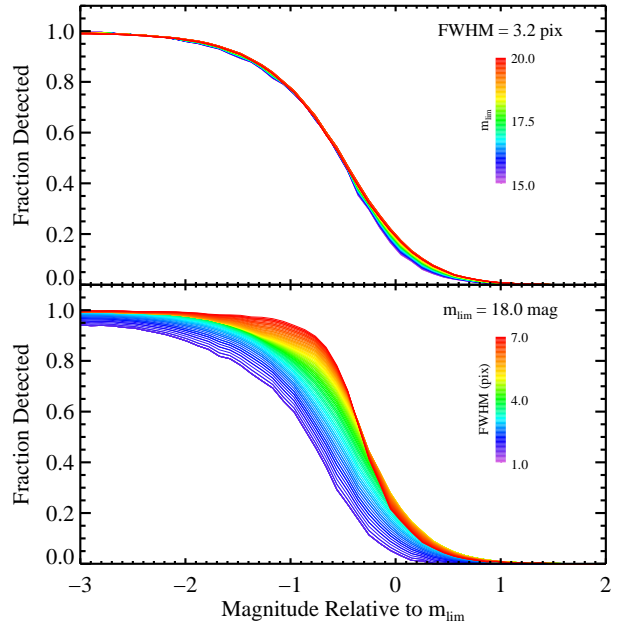}
\caption{ Detection efficiency for simulated point sources as a
  function of magnitude relative to the logged limiting magnitude,
  $m_{\rm lim}$. The top panel shows that the shape of the detection
  efficiency curve is nearly independent of the limiting magnitude for
  a fixed FWHM. The lower panel shows that for a fixed $m_{\rm lim}$,
  the shape of the detection efficiency curve depends strongly on the
  average FWHM of the image. For both plots, the 50\% completeness is
  brighter than the logged limit mainly because the simulations
  require a $5\sigma$ detection while the logged limits are derived
  from all extracted sources (i.e. $3\sigma$ detections are included).
}
\label{fig:de_curves}
\end{figure}

Figure~\ref{fig:de_curves} shows the average $5\sigma$ detection
efficiency curves for isolated objects. As can be seen, the shape of
the curve is robust to changes in the limiting magnitude, but the
effect from variations in the FWHM is quite noticeable.

Accurate limiting magnitudes were not recorded for the co-added images
during the first half of our survey, but we do have the limits for the
individual images. Using our later data, we determine an empirical
relation between the limits of the individual images and the coadds
derived from them. We started by assuming the limiting magnitudes for
each individual image is approximately equal to three times the sky
noise in the aperture. Inverting this, we can find the sky noise
expected on the co-added frame from the sum of the individual images,
and turn this into a limiting magnitude guess. Comparing these guesses
to the actual values recorded in the logs over the second half of our
survey, we find that the agreement is good for the deepest data, but
there is a systematic offset for more shallow coadds. We perform a
linear fit to these data and use it to derive magnitude limits for the
coadds created in the beginning of the survey.

\section{Survey Efficiency}\label{survey_eff}

It is not enough to merely detect a source to count it as a discovery;
each such candidate must pass additional screening to be included in
our sample. This includes both machine cuts designed to reject false
positives and cuts made by the humans who have the final say in
elevating candidates to discoveries. Often it is straight forward to
determine the effects of the machine cuts through a Monte Carlo
process: simulated supernovae drawn from the expected population are
``observed'' using the actual survey cadence and limiting magnitudes,
and the resulting ``candidates'' are passed through the same machine
vetting process as the real data. The sensitivity of the survey to
sources with various apparent brightnesses and light curves can then
be directly determined by noting the fraction of the simulated
supernova that pass. The effects humans play in the ultimate selection
of candidates could in principle be determined through a similar
simulation if the test were performed concurrently with the actual
survey to account for learned behavior. To get statistically
meaningful results, the human scanner would need to be presented with
an enormous volume of simulated data, which is far too large of a
burden on the researchers to be done in practice, although previous
studies have performed limited versions of such human experimentation,
which guide our tests.

As stated above, the basic machine cuts applied to the raw candidates
are 1) the FWHM should be between 0.5 and 2 times the image average,
and 2) the signal-to-noise should be at least $2.5\sigma$ on SUB1 and
SUB2, and $5\sigma$ for SUB. To help reduce the number of false
positives created by imperfect subtraction of galaxy light, we also
required that candidates near the cores of cataloged galaxies show an
increase of at least 5\% in flux over the reference template (given
the bright limits of our survey and prior coverage of our search
fields by galaxy surveys, it is fair to assume that almost all
galaxies as bright or brighter than our \SNeIa\ sample were
cataloged). We are mainly sensitive to \SNeIa\ brighter than $-19.0$
absolute, so hosts fainter than about $-20.3$ are unaffected by this
cut. Of course, when the host galaxy is resolved, only the fraction of
light coincident with the candidate matters in this percentage
increase cut. For example, we recorded a 50\% increase for SN~2005bg
($M_{\rm peak} = -19$), which was discovered near the core of a
$M_{r}\sim-21$ host. Candidates coincident with sources not listed in
the galaxy catalogs (typically foreground stars) were required to show
a 15\% increase. The remaining candidates located more toward the
outskirts of cataloged galaxies or in isolated regions were not
required to show a minimum increase in flux. We do not include this
selection criteria in our Monte Carlo simulations, so our efficiency
in discovering supernovae in luminous host galaxies may be
overestimated, which again means that our rate in giants may be
underestimated.

The basic setup for our efficiency calculation is similar to that used
in previous rate studies \citep[cf.][]{dilday2008}. We determine the
efficiency as a function of luminosity distance by considering a
series of thin shells over which the efficiency (and rate) can be
assumed to be constant. In each of these shells we simulate a number
of \SNeIa\ randomly drawn from the pseudo-absolute magnitude
distributions discussed in \S\ref{LF}, choose an appropriate light
curve template and scale it to the given distance, and select a random
date for maximum light. By comparing the expected magnitudes on the
dates each simulated supernova would have been observed by our survey
and using the actual survey detection limits on those nights, we
determine the fraction of simulated \SNeIa\ that could have been
detected in each distance bin. We take into account the detection
efficiency curve discussed in \S\ref{DE} appropriate to a given
observation based on the limiting magnitude and FWHM recorded (or
estimated). For each observation epoch of each simulated supernova, we
draw random numbers and compare these to the detection probabilities
to determine if a simulated detection is made on each of the SUB1,
SUB2, and SUB. For each of our 177 fields, we simulated 100 \SNeIa\ in
each of 360 distance shells spaced logarithmically from 40 to
1000\,Mpc, which totals to over six million simulations for each pAMD
assumed. To naturally account for the changing number of survey
fields, we allow simulated supernovae to peak anytime in any field
during the entire survey period including off season.

\begin{figure*}
\includegraphics[width=\linewidth]{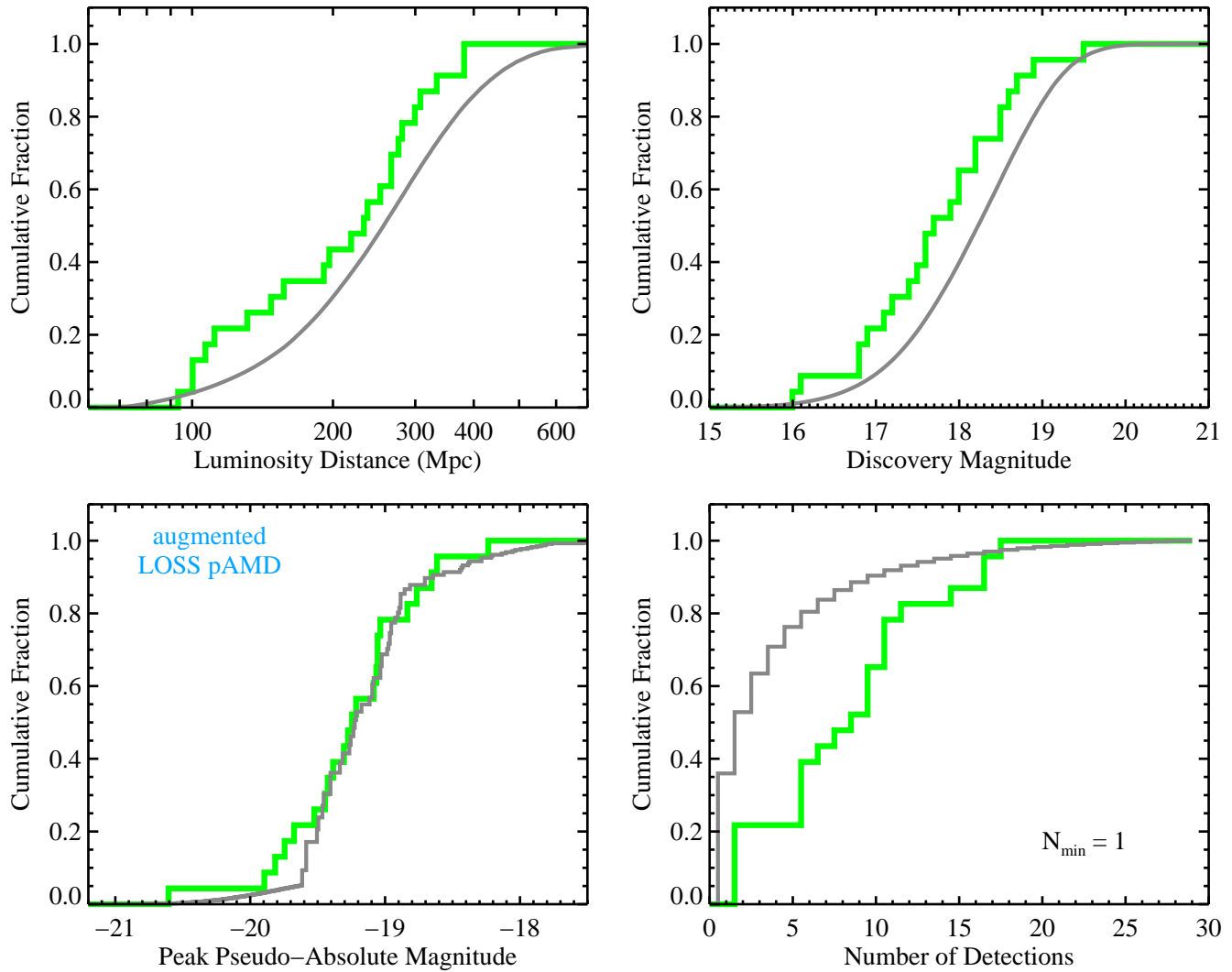}
\caption{ Distributions predicted from the Monte Carlo simulations
  (grey) compared to the observed sample (green) before correction of
  selection bias. The augmented LOSS pseudo-absolute magnitude
  distribution (see \S\ref{lossLF}) is assumed \citep{li2011a}.}
\label{fig:initial_cdf}
\end{figure*}

To assess how accurately these simulations model the actual survey, we
compare the luminosity distance, discovery magnitude, peak absolute
magnitude, and detection number distributions predicted by the Monte
Carlo simulations to the observed distributions.
Figure~\ref{fig:initial_cdf} shows that the survey model discussed
thus far is a poor representation of the true sample if the augmented
LOSS pAMD is adopted (the same result holds if the SDSS-II model pAMD
is used). The model predicts the \SNeIa\ should be further away than
observed, discovered at magnitudes fainter than observed, and the
simulated supernovae are typically detected on fewer nights. This last
result is key to understanding the failure of the model thus far. The
model predicts that $>35$\% of the \SNeIa\ should be detected by
the automatic search selection cuts on exactly one night. After all,
the largest volume element is for the most distant events that are
only detectable at their peak. Since we have zero \SNeIa\ detected on
just one night, it would seem that our survey is biased against such
``one-nighters.''  A possible explanation for this is the influence of
the human scanners who have the final say in selecting
targets. Although it was never a formal requirement of the search that
events be detected on more than one night (and in fact, we triggered
spectroscopic follow-up more than once on the same night based on a
first detection), the informal requirement that candidates ``look
good'' may have created a bias against the weakest detections. Indeed,
\citet{dilday2008} studied the response of human scanners to simulated
supernovae and determined that there is such a bias against
one-nighters.

\begin{figure*}
\includegraphics[width=\linewidth]{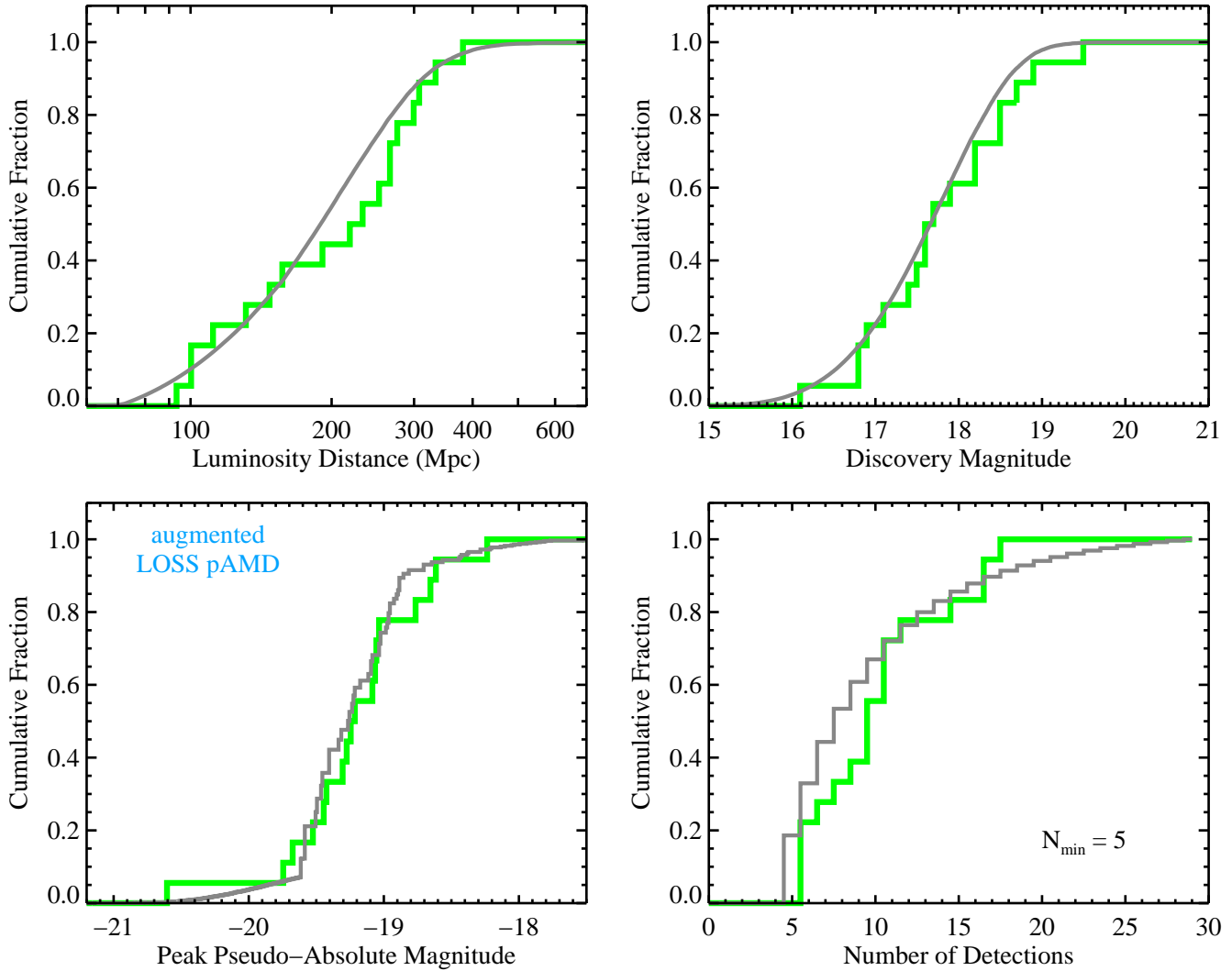}
\caption{ Distributions predicted from the Monte Carlo simulations
  (grey) compared to the observed sample (green) after correction for
  selection bias. The augmented LOSS pseudo-absolute magnitude
  distribution (see \S\ref{lossLF}) is assumed \citep{li2011a}.}
\label{fig:final_cdf}
\end{figure*}

To account for our apparent selection bias against weak detections, we
apply a cut to the simulations and the real sample alike on the number
of nights detected. We find that setting a minimum number of 5 nights,
the model and actual distributions agree for both the augmented LOSS
pAMD and the SDSS-II model. This cut preferentially removes the most
distant supernovae with the faintest observed magnitudes, so after
applying this single correction, all of the model distributions
considered agree with the actual observed sample
(Fig.~\ref{fig:final_cdf}). The minimum number of detections was set
by calculating the probability that the observed sample was drawn from
the model distribution via a KS test. We assign a consistency
probability, $P_{KS}$, by randomly drawing from the model distribution
and comparing the maximum displacement in the cumulative distributions
of the model and each draw. For 4 or fewer detections and the LOSS
pAMD, only 0.3\% (or 0.4\% with the SDSS-II pAMD model) or less of the
random draws have a displacement as large or larger than the real data
compared to the models, so we reject these models. With a minimum of 5
detections, the probability rises to 9\% (5\%), so we cannot reject
the null hypothesis that the data were drawn from this model
distribution. In this case, the probabilities for the other three
distributions are considerably higher, so the model does not appear to
be inconsistent with the data.

We also tested the effects of systematically shifting the input pAMDs
in brightness. If we shift the augmented LOSS pAMD 0.3\,mag fainter,
both the predicted distance and absolute magnitude distributions
become incompatible with the observed distributions (as above we
require pipeline detections on 5 or more nights). Shifting the LOSS
pAMD 0.2\,mag brighter, the predicted and observed distance
distributions fall into agreement, but the absolute magnitude
distributions are not compatible. With the SDSS-II pAMD model, a shift
of either 0.2\,mag fainter or brighter can be ruled out. This
demonstrates that we would be able to detect if the magnitude system
of the pAMD employed was systematically biased with respect to our
unfiltered magnitudes by more that expected from our checks in
\S\ref{LF}. We will use the allowed magnitude offsets to estimate our
systematic error in \S\ref{conclusions}.

\begin{figure}
\includegraphics[width=\linewidth]{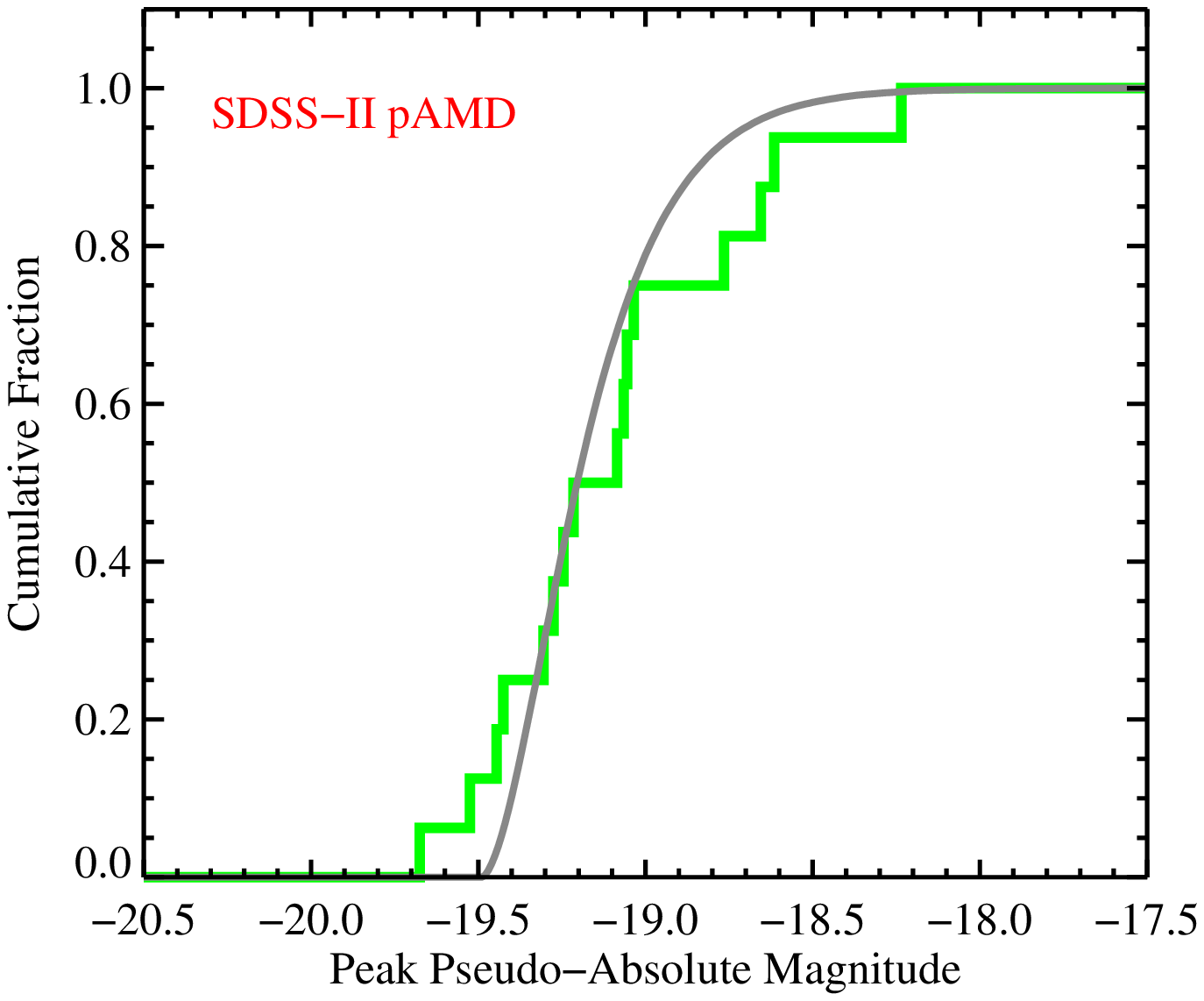}
\caption{ Peak pseudo-absolute magnitude distribution recovered from
  the Monte Carlo simulations assuming the SDSS-II model magnitudes
  (grey; see \S\ref{sdssLF}) compared to the observed sample (green)
  after correction for selection bias. }
\label{fig:sdss_model}
\end{figure}

\begin{figure}
\includegraphics[width=\linewidth]{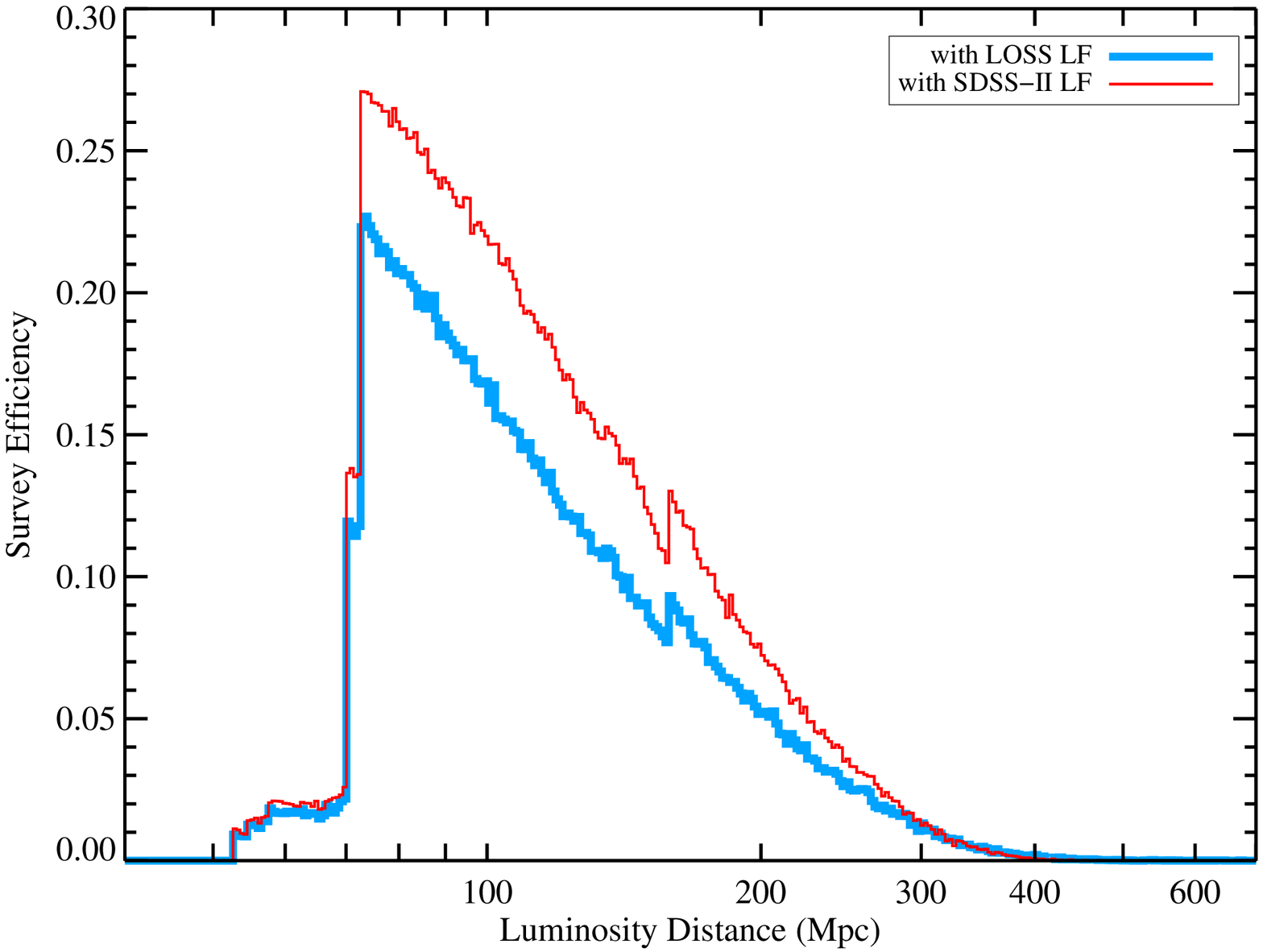}
\caption{ Search efficiency assuming the modified LOSS pseudo-absolute
  magnitude distribution (blue) or the SDSS-II model (red) and
  averaged over all fields and the full survey time-span (including
  un-searched periods). The sudden jumps in efficiency mark the minimum
  allowable distances for various fields (e.g. targets in the Virgo
  fields must be at least 72\,Mpc away to meet our selection cuts).  }
\label{fig:SE}
\end{figure}

\section{Host Galaxies}\label{hosts}

\citet{li2011b} found that lower luminosity galaxies produced more
\SNeIa\ per unit luminosity than high luminosity hosts. In order to
compare the host luminosity distribution of our sample to the SDSS-II
and the expectations of the LOSS luminosity specific rates coupled
with a galaxy luminosity function, we have measured ugriz photometry
for each supernova host. We make deep stacks from the SDSS DR7 data
\citep{sdss_dr7} in each of the 5 bands, excluding data obtained while
the supernovae were active (assumed to be $-20$\,d$ < t_{max} <
100$\,d in the rest frame), and retain the median using a slightly
modified version of the Montage package\footnote{
  \url{http://montage.ipac.caltech.edu/}}. Cuts on the sky background
and PSF size are used to reject the worst images (about a third of the
available data). We identify the host galaxies following
\citet{sullivan2006}. Briefly, we use the SExtractor shape parameters
($C_{xx}$, $C_{xy}$, and $C_{yy}$) to define an ellipse for each
object in the field, and we identify the host as the object that
includes the supernova position inside its ellipse with the smallest
possible scaling unless the scaling factor, $R$, is greater than 5. If
this is the case, we perform forced photometry at the location of the
supernova in a $3''$ diameter aperture. Otherwise we run SExtractor in
dual image mode to measure the host galaxy photometry in consistent
aperture sizes on the best matching host. We choose to perform the
object detection and aperture definition on the r-band images.

A few host galaxies are worthy of special mention: {\bf 1)} The host
of SN~2007if is not detected in the co-added SDSS data. We instead use
the g-band magnitude of the host reported by \citet{childress2011} to
scale a spectrum of the host galaxy and then measure synthetic
photometry in the SDSS bands from this. {\bf 2)} The LOSS team
assigned the giant galaxy KUG\,1259+286 as the host of SN~2005ck, but
the SDSS redshift for this galaxy is incompatible with the supernova
\citep{pugh2005, leaman2011}. There is a small galaxy nearby,
SDSS\,J130218.65+282046.8, for which we find a spectroscopic redshift
consistent with SN~2005ck; however, this galaxy is slightly too far
away to meet the host selection criteria described above (as would be
KUG\,1259+286), and it is therefore considered hostless. Identified or
not, the host must be a dwarf, so this should not affect our
conclusions. {\bf 3)} SN~2007sp is located a considerable distance
($R>10$) from a large galaxy at a similar SDSS derived
redshift. Although this may yet be the host, for consistent
application of the \citet{sullivan2006} selection method, we must
consider this target to be hostless. {\bf 4)} There is a galaxy
coincident with the location of SN~2007kh, but spectroscopic follow-up
shows this to be a $z\sim0.5$ galaxy, which is well in the
background. The true host of SN~2007kh is therefore a mystery, so we
simply adopt the photometry of the background source, assume the
redshift of SN~2007kh, and take this as an upper limit on the host
luminosity. {\bf 5)} The location of SN~2007op is outside of the SDSS
DR7 footprint; however, the host is included in DR8, so we adopt the
photometry from the DR8 pipeline \citep{aihara2011}.

In addition, we perform the same analysis on the full set of 516
spectroscopically confirmed, probable, or photometrically probable
$z<0.3$ \SNeIa\ from the SDSS-II \citealp{dilday2010}. The larger
number of visits to Stripe82 over the course of the SDSS-II results in
co-added images (again excluding those contaminated by supernova
light) with limiting magnitudes considerably deeper than the search
itself. As a result, even $M\sim-17$ dwarf galaxies can be detected
out to the $z\sim0.3$ limit of the sample. As with the ROTSE-IIIb
sample, we manually reviewed the selection of each host galaxy and
noted any special circumstances that might produce misleading results
(the potential error rate appears similar for the ROTSE-IIIb and
SDSS-II samples). As needed, we adjusted the SExtractor parameter {\tt
  DEBLEND\_MINCONT} to either ensure large galaxies were not
inadvertently split into smaller pieces or to attempt to split a
likely galaxy blend. The latter issue was not always resolved, both
for the ROTSE-IIIb and SDSS-II hosts. Of particular note, a limited
number of SDSS-II hosts appear to be dwarf galaxies heavily blended
with (if not merging into) larger galaxies, such as SDSS-II
SN03592. When such blends are not separated by SExtractor, we perform
forced aperture photometry at the location of these supernovae to
better reflect the luminosity of the dwarf hosts.

Since the host galaxies are distributed over a range of redshifts, we
must perform K-corrections to compare the galaxy luminosities in the
same rest-frame band passes. This is accomplished using the {\tt
  kcorrect.pro} package, version 4.2 \citep{blanton2007}. For
comparison against the literature, we calculate absolute B- and K-band
magnitudes for our hosts in the Vega system. These are recovered from
the best fit galaxy template through {\tt kcorrect.pro}. We compared
our derived K-band magnitudes (based only on the optical SDSS
photometry) for a few low-redshift hosts to detections in the 2MASS
survey \citep{skrutskie2006}, and found good agreement.

\begin{figure*}
\includegraphics[width=\linewidth]{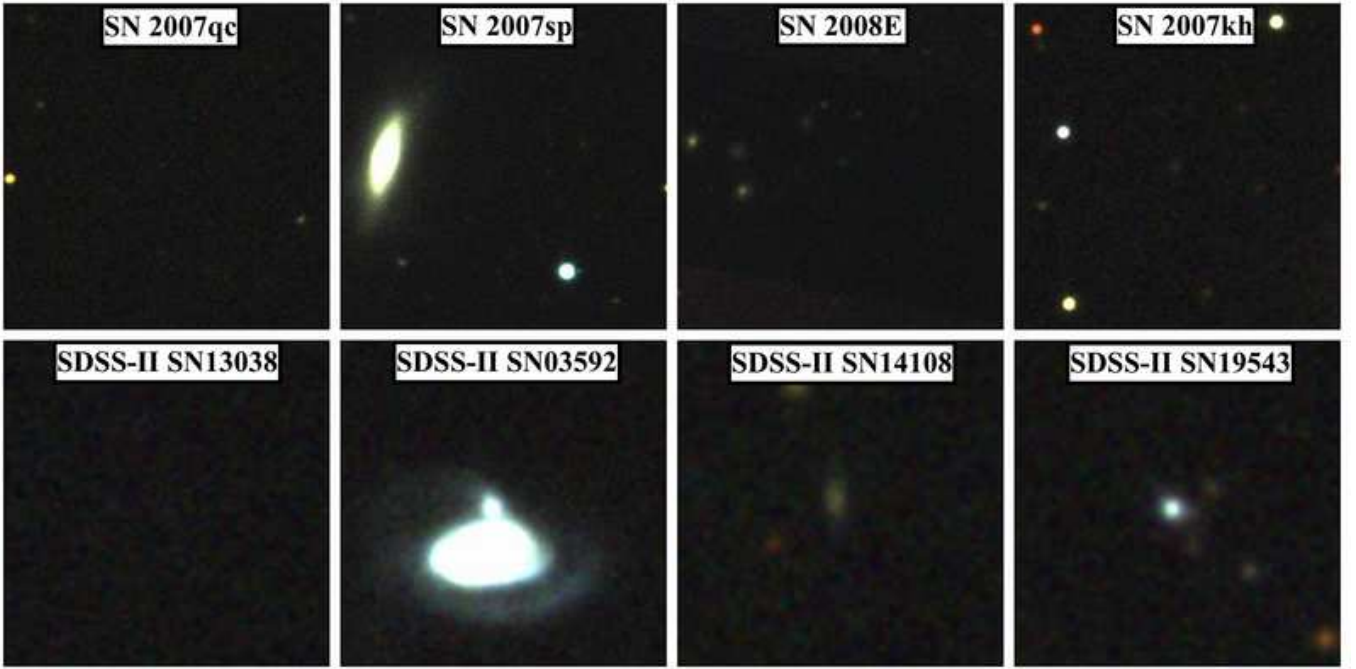}
\caption{ The four lowest luminosity hosts from the ROTSE-IIIb
  \SNeIa\ sample (top row) compared to the four faintest hosts from
  the SDSS-II $z<0.15$ sample (bottom row; \citealt{dilday2010}). Each
  panel shows a $30 \times 30$\,kpc close-up centered on the SN
  position made from SDSS images. SN~2007sp is offset from a luminous
  host at the same redshift, but this separation is more than double
  that allowed by the \cite{sullivan2010} host selection criteria, so
  we must consider the host undetected. SDSS-II SN03592 may be hosted
  by a dwarf galaxy merging with a larger host. We were unable to
  separate these galaxies with SExtractor, so we instead perform
  forced aperture photometry, which may underestimate the brightness
  of the dwarf. A faint galaxy is seen at the position of SN~2007kh,
  but spectroscopy indicates this is a background source, so we take
  its measured brightness as an upper limit on the actual host. }
\label{fig:faint_hosts}
\end{figure*}

\begin{figure*}
\includegraphics[width=\linewidth]{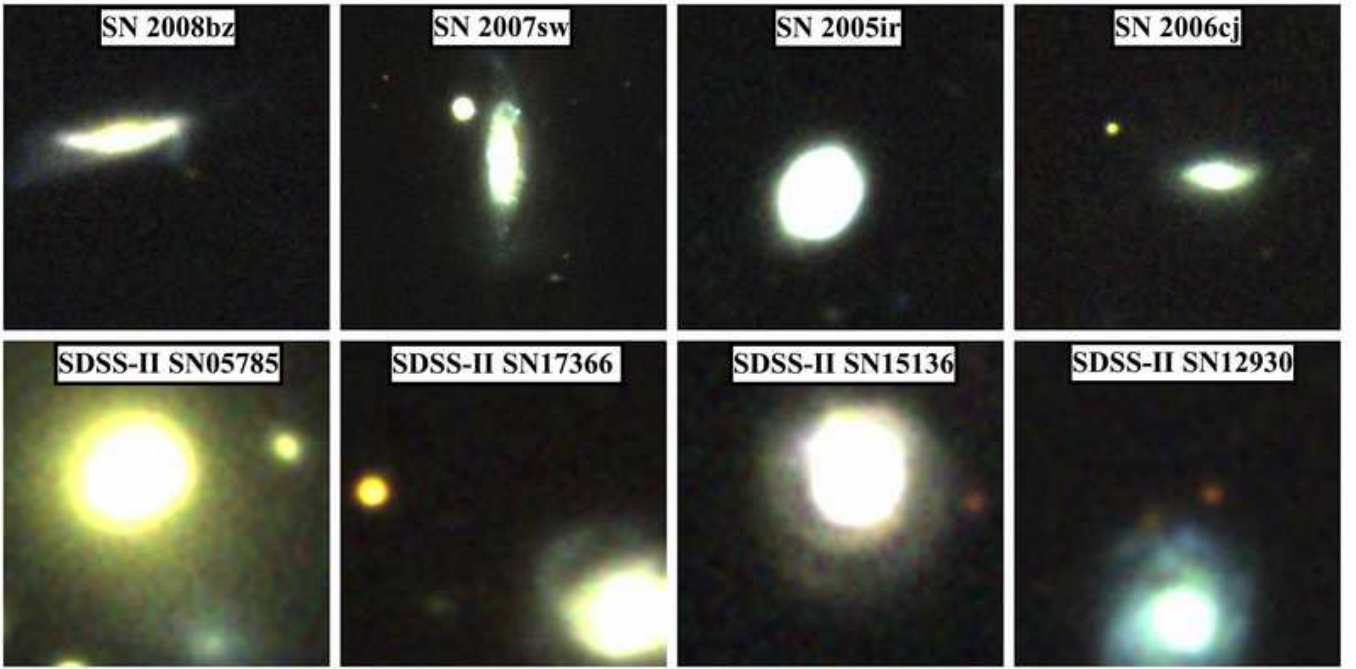}
\caption{ Similar to Figure~\ref{fig:faint_hosts} but for the four
  highest luminosity hosts in the ROTSE-IIIb sample (top row) and the
  SDSS-II (bottom row).}
\label{fig:bright_hosts}
\end{figure*}

\begin{figure}
\includegraphics[width=\linewidth]{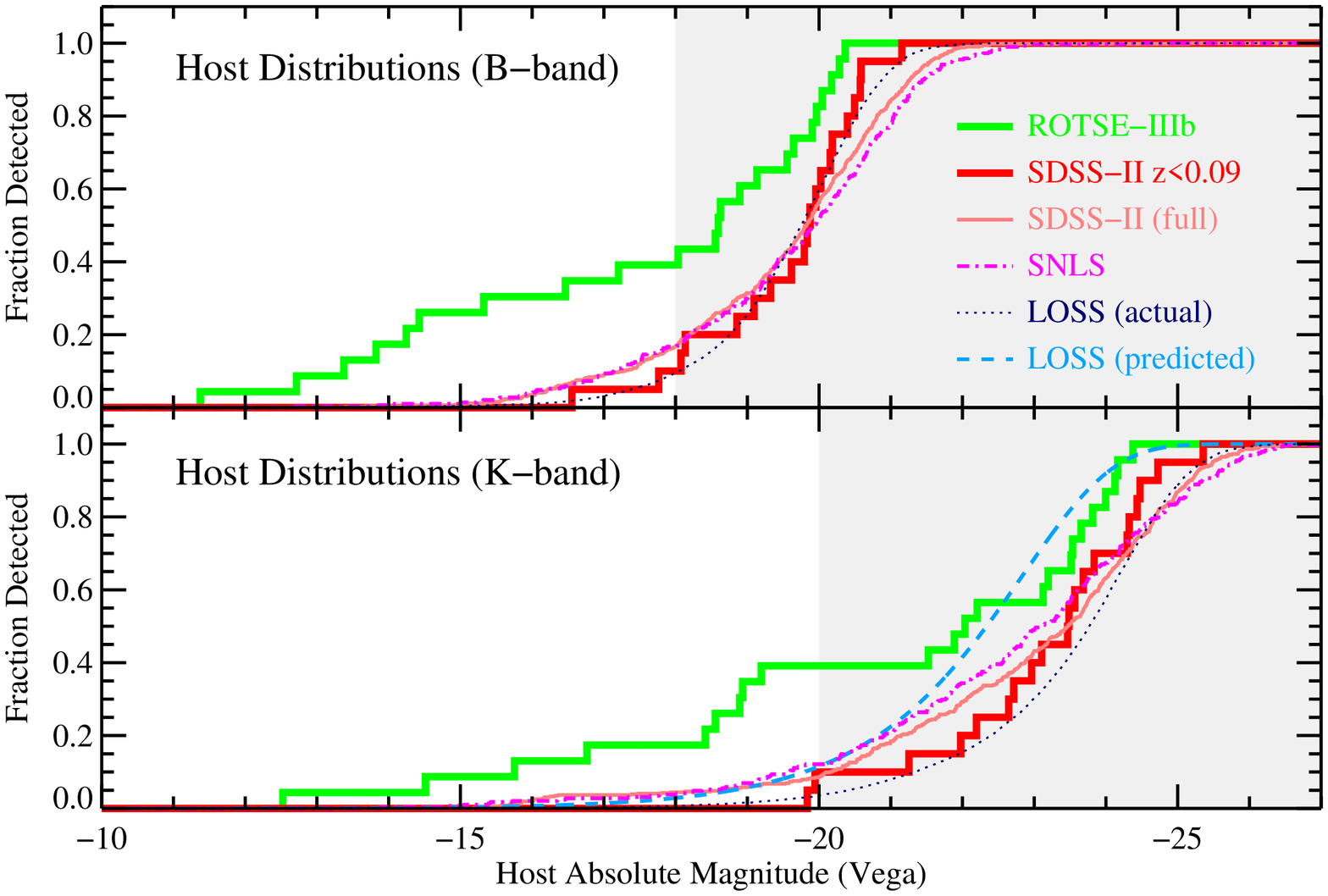}
\caption{ Absolute magnitude distributions for the host galaxies of
  \SNeIa. A few of the hosts fainter than $M_B \simgt -17$ were not
  detected. We employ upper limits for these events, but this does not
  affect our results. The host galaxies of the ROTSE-IIIb supernovae
  (green line) are typically fainter that hosts from the SDSS-II
  search in a similar redshift range (red line), and the ROTSE-IIIb
  sample includes a larger fraction of dwarf hosts than seen or
  predicted through the other surveys shown. }
\label{fig:hosts}
\end{figure}

The four lowest luminosity hosts from the ROTSE-IIIb and nearby
SDSS-II samples are shown in Figure~\ref{fig:faint_hosts} and the four
highest luminosity hosts from each sample are shown in
Figure~\ref{fig:bright_hosts}. In Figure~\ref{fig:hosts}, we compare
the host galaxies of the ROTSE-IIIb sample to the 20 hosts from the
SDSS-II sample with redshifts $z<0.09$ \citep{dilday2010}. These two
samples cover a similar redshift span and can thus be compared
directly. As can be seen, our sample includes relatively few
high-luminosity host galaxies and, of particular interest, a number of
very low luminosity hosts. A KS test gives only a 4\% chance that
these ROTSE-IIIb and low redshift SDSS-II samples are drawn from the
same population. If we assign SDSS-II SN03592 to the larger host, this
probability drops to just 1\%. The agreement likelihood with the full
population of galaxies hosting the \SNeIa\ from the SDSS-II rate
calculation is about 0.2\%. For this larger sample, a small fraction
of cases that have undetected hosts also have limiting magnitudes
brighter than detected ROTSE-IIIb hosts, so the exact distribution
below $M_B \sim -17$ is uncertain. However, this does not affect our
key result: 43\% of the ROTSE-IIIb hosts are $M_B>-18$\,mag dwarf
galaxies while only 17\% of the full SDSS-II and 2 out of 20 from the
$z<0.09$ sample are hosted by such dwarfs (including SDSS-II SN03592).

Figure~\ref{fig:hosts} also shows the host galaxy distributions from
the SNLS \citep{astier2006}. As with the SDSS data, we measure the
host galaxy photometry using SExtractor in dual image mode, this time
using the i-band as reference, and we determine the rest frame
absolute magnitudes with {\tt kcorrect.pro}. We use the MegaPipe
median image stacks provided through CADC \citep{gwyn2008}. To check
if these measurements are biased by the inclusion of images with the
supernovae detected in the determination of the medians, we compare
our i-band observer frame measurements to
\citet{sullivan2010}. Surprisingly, we find that our measurements are
actually systematically fainter than the \citet{sullivan2010} values
by $\sim0.3$\,mag. For example, for the host of SNLS~03D1ar, we find
$m_i=19.84$ without correcting for 0.05\,mag of Galactic reddening
\citep{sfd1998}. This agrees with the value in the MegaPipe catalog,
and it is consistent with the SDSS DR8 ($19.88\pm0.04$), but it is
fainter than reported in \citet{sullivan2010} who find
$m_i=19.57$. Aside from this systematic offset, we do not find any
evidence that our photometry is biased by the supernova light even in
the faintest hosts. It is interesting that the B-band distributions of
the SDSS-II and SNLS host galaxies appear to diverge toward the bright
end, but there is a 12\% chance that a random draw from the SDSS-II
population would lead to an equal or larger displacement. As with the
SDSS-II, the deep coadds are sufficiently deep to detect all SNLS
hosts brighter than about $M_B \sim -17$, and again the ROTSE-IIIb
\SNeIa\ prefer fainter hosts.

Although LOSS targets specific galaxies, which biases the sample
against low luminosity hosts, they adopt the \citet{kochanek2001}
K-band galaxy luminosity function and factor in their rate-size
relation to determine the volumetric \SNeIa\ rate \citep{li2011b}. We
include the K-band host galaxy distribution expected by the LOSS rate
calculation in the lower panel of Figure~\ref{fig:hosts}. Clearly,
this expected distribution is different than the observed SDSS-II and
SNLS distributions. Specifically, LOSS assumes that 80\% of \SNeIa\ in
a given volume should come from hosts fainter than about
$K>-23.4$\,mag, but the SDSS-II finds only about 40\% of its $z<0.09$
hosts are fainter than this luminosity. This result holds
qualitatively even if the correction for the rate-size relation is
removed; among the giants, the SDSS-II sample shows a preference for
the higher luminosity hosts. Moving to lower luminosities, the
ROTSE-IIIb sample shows a considerably higher fraction in dwarf
galaxies.

As explained in \S\ref{survey_eff}, our selection process for
ROTSE-IIIb may have a slight bias against high-luminosity hosts. In
the next section we calculate the volumetric rates of \SNeIa\ in
dwarfs and giants for comparison against other samples. For
convenience, we will label galaxies fainter than $M_B>-18$ as
``dwarfs,'' and more luminous galaxies as ``giants.'' When discussing
the K-band magnitudes of the hosts, we will consider $M_K>-20$
galaxies to be dwarfs, since the $B-K$ colors for many of our
$M_B\sim-18$ hosts are around 2\,mag. If our sample's apparent
preference of low luminosity hosts is a result of a selection effect,
then our \SNeIa\ rate in giants should be lower than the actual rate
while the rate in dwarfs should be accurate. We compare our values to
published rates as a check.

\section{Rates}\label{rates}

In this section we calculate the volumetric \SNeIa\ rate,
$\mathcal{R}$, using: 

\begin{equation}
\mathcal{R} = \frac{N_{\rm obs}}{\epsilon V t}
\end{equation} \label{eqn:rate}

\noindent where $N_{\rm obs}$ is the observed number of events, $t$ is
the {\it proper} time of the survey, $V$ is the co-moving volume
surveyed, and $\epsilon$ is the efficiency factor that gives the
estimated fraction of events actually discovered to the true number of
events in the time and volume covered (i.e. $\epsilon \leq 1$).

We have 18 \SNeIa\ appropriate to the augmented LOSS pAMD and 16
appropriate for the SDSS-II model that were detected on 5 or more
nights (see \S\ref{LF} and \ref{survey_eff}). We calculated the
denominator of equation \ref{eqn:rate} in \S\ref{survey_eff} as a sum
over a series of logarithmically spaced distance bins to account for
the decline in survey efficiency with distance (see figure
\ref{fig:SE}). The proper time for the survey is calculated by
correcting the observer frame span for time dilation in each
luminosity distance bin. The survey period was set from November 1,
2004 through January 31, 2009, with the former date reflecting the
month the survey began and the latter set to the end of the standard
2008B astronomical semester. The final term for the survey Volume is
calculated for each distance bin by integrating the co-moving volume
element of the Friedmann-Robertson-Walker metric with our chosen
cosmological parameters (a flat, $H_0=71$\,km\,s$^{-1}$\,Mpc$^{-1}$,
$\Omega_m=0.27$ universe) and factoring in the fraction of sky covered
by our survey.

Special care must be taken to determine the actual survey area since
the galaxy clusters of the survey are covered by an irregular grid of
overlapping fields (see Figs. \ref{fig:winter_spring} and
\ref{fig:summer_fall}). The typical pointing error of ROTSE-IIIb is
about $\pm 0.05$ degrees, so we assume the reference templates, which
represent the intersection of numerous individual images, cover $1.75
\times 1.75$ degrees each. The actual pitch for the main search fields
is 1.65 degrees, so this is the effective size for fields bordered on
all sides. To determine the total survey area, we construct an all sky
register similar to the images in Figures \ref{fig:winter_spring} and
\ref{fig:summer_fall}, and then total all the sky area visited at
least once during the survey period. Our total sky coverage is 499.0
square degrees.

\begin{figure}
\includegraphics[width=\linewidth]{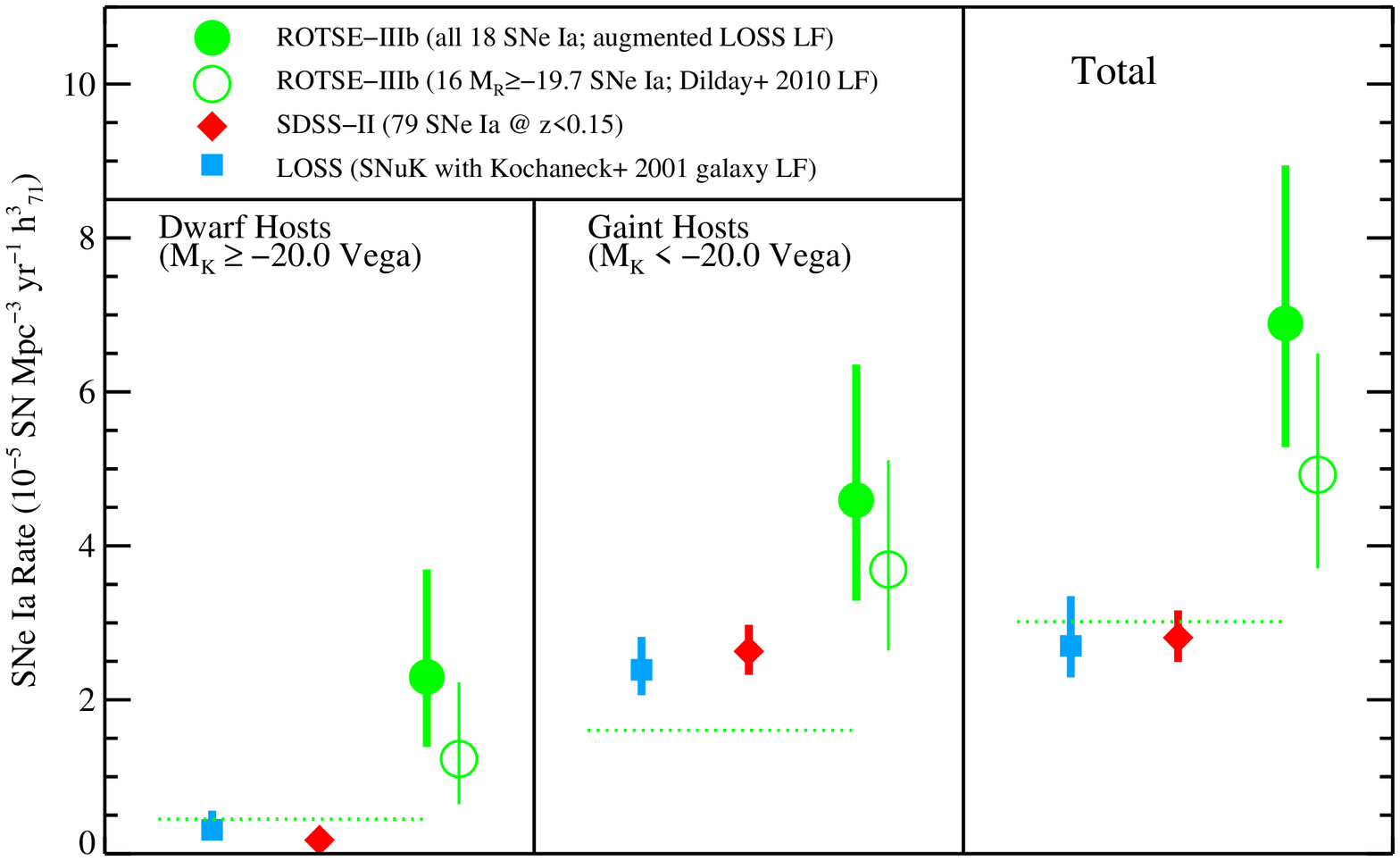}
\caption{ Volumetric \SNeIa\ rates. The right panel shows the total
  volumetric rate from ROTSE-IIIb (green circles; this work), SDSS-II
  (red diamond; \citealt{dilday2010}), and the LOSS SNuK rate combined
  with a K-band galaxy luminosity function (blue square;
  \citealt{li2011b,kochanek2001}). The vertical bars show the
  statistical error only. The panels to the left show the
  contributions to the total volumetric rate from dwarf and giant
  hosts. The dotted horizontal lines mark the $3\sigma$ lower limits
  for the complete ROTSE-IIIb samples in each group. The filled green
  circle shows the ROTSE-IIIb rates computed assuming the LOSS
  \SNeIa\ pseudo-absolute magnitude distribution \citep{li2011a}, and
  the open circles show the values derived using the same
  \SNeIa\ distribution assumed for the SDSS-II calculation
  \citep{dilday2008, dilday2010} and removing spectroscopically
  confirmed \SNeIa\ in a similar manner to the SDSS-II vetting
  process. }
\label{fig:rates}
\end{figure}

Adopting the LOSS \SNeIa\ pseudo-absolute magnitude distribution with
a supplemental population of high luminosity events (assumed to be 1\%
of the total population; see \S\ref{lossLF}) we find a rate of
$(6.9^{+2.1}_{-1.6})\times10^{-5}$\,\SNeIa\,Mpc$^{-3}$\,yr$^{-1}$\,$h_{71}^{3}$
(statistical error only). With the SDSS-II model pAMD, the volumetric
rate drops to the lower (but statistically consistent) value of
$(4.9^{+1.6}_{-1.2})\times10^{-5}$\,\SNeIa\,Mpc$^{-3}$\,yr$^{-1}$\,$h_{71}^{3}$.
These total rates and the rates specific to dwarf and giant hosts are
shown in Figure~\ref{fig:rates} and listed in table~\ref{table:rates}.

\begin{deluxetable*}{lcccccc}
\tablewidth{0pt}
\tablecaption{Type~Ia Supernova Rates}
\tabletypesize{\scriptsize}

\tablehead{
  \colhead{Survey} &
  \colhead{$N_{\rm Ia, dwarfs}$} &
  \colhead{Dwarfs} &
  \colhead{$N_{\rm Ia, giants}$} &
  \colhead{Giants} &
  \colhead{$N_{\rm Ia, total}$} &
  \colhead{Total} 
}
\startdata
LOSS & \nodata & $
0.31^{+0.25}_{-0.10}$ & \nodata & $2.39^{+0.43}_{-0.33}$ & \nodata & $2.70^{+0.65}_{-0.41}$  \\
SDSS-II ($z<0.15$) & 5 & $0.18^{+0.12}_{-0.08}$ & 74 & $2.63^{+0.34}_{-0.30}$ & 79 & $2.81^{+0.35}_{-0.32}$  \\
{\bf ROTSE-IIIb} (w/ LOSS LF) & 6 & $2.31^{+1.41}_{-0.92}$ & 12 & $4.63^{+1.78}_{-1.32}$ & 18 & $6.94^{+2.07}_{-1.62}$  \\
{\bf ROTSE-IIIb} (w/ SDSS-II LF) & 4 & $1.23^{+1.00}_{-0.59}$ & 12 & $3.69^{+1.42}_{-1.05}$ & 16 & $4.92^{+1.58}_{-1.22}$  \\
\enddata

\enddata
\tablecomments{
Rates in units of $10^{-5}$\,\SNeIa\,Mpc$^{-3}$\,yr$^{-1}$\,$h_{71}^{3}$
}
\label{table:rates}
\end{deluxetable*}

For comparison, the SDSS-II found a \SNeIa\ rate of
$(2.81^{+0.35}_{-0.32})\times10^{-5}$\,\SNeIa\,Mpc$^{-3}$\,yr$^{-1}$\,$h_{71}^{3}$
using their $z<0.15$ \SNeIa\ and assuming the rate to be constant out
to this limit \citep{dilday2010}. Since the \SNeIa\ rate increases
with redshift and even the lowest SDSS-II redshift bin extends out to
larger distances than the ROTSE-IIIb sample, the SDSS-II rate may be
slightly inflated with respect to our more nearby rate measurement. In
any case, the SDSS-II rate is below our $3\sigma$ lower limit derived
with the augmented LOSS pAMD and may be incompatible with the SDSS-II
model pAMD as well at about the $2\sigma$ level. We split the samples
into dwarf and giant hosts and find that while the ROTSE-IIIb rates in
giants are marginally consistent with the SDSS-II results, our rate in
dwarfs is significantly higher than the SDSS-II with either the
augmented LOSS or SDSS-II model pAMD.

To convert the LOSS host luminosity specific \SNeIa\ rate into a
volumetric rate, we follow similar steps as reported in
\citet{li2011b}, but with an important difference. We start with the
rate-size relation and SNuK measurement from LOSS and combine this
with the local galaxy K-band luminosity function
\citet{kochanek2001}. Like the LOSS estimate, the calculation is done
separately for early- and late-type galaxies, and then combined to
give the final result. The published LOSS rate is determined by
multiplying the local galaxy K-band luminosity densities ($j_{\rm
  early}$ and $j_{\rm late}$) by average values of their SNuK, which
are determined from a weighted average involving the number densities
from the galaxy luminosity functions. We choose to recalculate the
LOSS rate through what we believe to be a more direct approach. We
simply multiply the galaxy luminosity function by the rate-size
relation function and integrate. With our approach, we find the
volumetric rate implied by the LOSS sample is
$(2.70^{+0.65}_{-0.41})\times10^{-5}$\,\SNeIa\,Mpc$^{-3}$\,yr$^{-1}$\,$h_{71}^{3}$. This
is consistent with, if not slightly lower than, the published LOSS
rate \citep{li2011b}. We compute the statistical error through a Monte
Carlo calculation including the uncertainty in the galaxy luminosity
function, which gives a somewhat larger confidence interval than
reported by LOSS. This rate is consistent with the SDSS-II rate, which
is an odd result given the apparently different \SNeIa\ populations
studied (see \S\ref{loss_v_sdss}). Thus, as was the case with the
SDSS-II rate, the total rates we derive from our ROTSE-IIIb sample
are higher at the $3\sigma$ level. We also determine the LOSS rate in
dwarfs and giants by simply changing the integration interval. Again,
LOSS and SDSS-II derive rates that are consistent with each other for
dwarfs, giants, and the total host population even though their
\SNeIa\ populations are apparently distinct (see further discussion in
\S\ref{conclusions}).

Using either \SNeIa\ pseudo-absolute magnitude distribution, our rate
in giants is comparable to SDSS-II and LOSS, so there is no evidence
that our selection process is biased against high luminosity
hosts. However, with the LOSS distribution our rate in dwarf hosts is
significantly higher than either SDSS-II or LOSS, and with the SDSS-II
model pAMD, our rate is higher at about the $2\sigma$ level.

\section{Conclusions}\label{conclusions}

We have used the background population of supernovae discovered by the
ROTSE-IIIb telescope over a four year period to derive the volumetric
rate of \SNeIa. Our rate is somewhat sensitive to the
\SNeIa\ pseudo-absolute magnitude distribution (pAMD) assumed in
deriving our overall search efficiency, and we find that recent rates
studies have used different, incompatible pAMDs. We have calculated
our rates using both the LOSS pAMD augmented with a small population
of high luminosity events (1\% of the total) and the same pAMD model
(and similar sample cuts) employed in the SDSS-II rate measurement to
quantify our distance dependent search efficiency. With the augmented
LOSS pAMD, we find a rate that is more than double published values in
a similar redshift range. This rate is higher than the SDSS-II or LOSS
rates at $99.9$\% confidence (ignoring systematics for the
moment). Similarly, performing the calculation on our
``normal-bright'' \SNeIa\ in a manner directly comparable to the
SDSS-II measurement (including the same pAMD model), our rate is still
nearly double that found by the SDSS-II out to a slightly higher
redshift. In this case, our rate is only higher than the SDSS-II value
at the $2\sigma$ confidence level, however.

The LOSS pAMD is utterly incompatible with the SDSS-II model, which,
in turn, is incompatible with the actual volume limited sample from
the first year SDSS-II. Neither the SDSS-II model nor the LOSS pAMD
allow for events as bright as SN\,2007if, so it would seem that none
of the available distributions account for the full population of
\SNeIa. Additionally, the large extinction value reported for the
photometrically selected SDSS-II SN09266 suggests that these
distributions may also fail to account for the true distribution of
host absorptions. Down to about -18.8\,mag, the LOSS and SDSS-II model
pAMDs agree, but the LOSS sample shows almost double the relative
contribution from fainter events. The solid agreement between the LOSS
and SDSS-II rates is perplexing given that they are apparently
sampling completely different \SNeIa\ populations.

The systematic errors in our rate measurement are dominated by the
uncertainty in the \SNeIa\ pAMD, including how such distributions
derived from (mostly) filtered photometry apply to our unfiltered data
set. Our total rate estimate with the SDSS-II model pAMD is 30\% lower
than our rate with the augmented LOSS pAMD. Part of this is due to the
exclusion of SNe 2007if and 2008ab from the SDSS-II model rate due to
their high luminosities (both of these events also have dwarf
hosts). The larger fraction of low-luminosity events with the LOSS
pAMD makes up the additional difference. Our search is mainly
sensitive to background supernovae with pseudo-absolute magnitudes
brighter than about $M_R = -18.5$\,mag, so at least 42\% of
\SNeIa\ are definitely missed by our search if the LOSS distribution
is assumed, but this fraction falls to just 8\% if we use the SDSS-II
model. We use our rate based on the SDSS-II model pAMD to constrain
the lower bound on the \SNeIa\ rate and our augmented LOSS rate for
the upper bound. Adding in a $\pm0.2$\,mag systematic offset to
account for the allowed uncertainty in our unfiltered magnitude system
with respected to the (filtered) pAMDs considered, our final
\SNeIa\ rate (including $1\sigma$ statistical errors) at a mean
redshift of $\overline{z} = 0.05$ is then between 3.7 and 13.3 in
units of $10^{-5}$\,\SNeIa\,Mpc$^{-3}$\,yr$^{-1}$\,$h_{71}^{3}$.

As our sample is spectroscopically complete, we do not include
uncertainty for unclassified candidates. This can be an important
concern for other, non-spectroscopically complete studies. For
example, the SDSS has reported\footnote{
  \url{http://sdssdp62.fnal.gov/sdsssn/snlist\_confirmed\_updated.php}}
561 spectroscopically confirmed \SNeIa, but there are at least 1070
photometrically probable \SNeIa\ uncovered by the same survey, and
presumably many more transient candidates in all including peculiar
\SNeIa, which may not pass the photometric screening
\citep{dilday2010, sako2011}.

There is also the question of whether all objects technically
classified as \SNeIa\ from their spectra should really be grouped in
the same physical category. For example, some have argued that some
peculiar \SNeIa, such as SN\,2005hk, may not be thermonuclear
supernovae \citep[cf.][]{valenti2009}. One event in our sample,
SN\,2006ct, bears some resemblance to this subclass, although we note
some differences as well in \S\ref{sample_notes}. Removing this one
event would lower our total rates by about 6\%. On the other hand,
some events grouped with CCSNe may actually be \SNeIa\ in disguise. In
particular, some Type\,IIn may be better suited in the \SNeIa\ camp
\citep{hamuy2003,dilday2012}. Over our survey period, we discovered
three background Type\,IIn supernovae (SNe\,2006db, 2006tf, and
2008am), but none of these have been linked to Type\,Ia explosions.

In using the augmented LOSS pAMD, we have attempted to measure the
total \SNeIa\ rate including Hubble diagram outliers and other
peculiar events. This should be directly comparable to the rate
measured by LOSS who include SN\,1991T-like, SN\,1991bg-like, and
SN\,2002cx-like events in their sample, but this has not always been
common practice in the literature. Many of the previous \SNeIa\ rate
measurements have explicitly excluded events that deviate from the
sub-sample of \SNeIa\ that are most useful as cosmological probes. For
example, the recent \SNeIa\ rates measured by the SNLS
\citep{perrett2012} are strictly valid only for \SNeIa\ with light
curve widths close to the nominal value ($0.8 < s < 1.3$, but see also
\citealt{gonzalez-gaitan2011}). \SNeIa\ like SN\,2003fg
(a.k.a. SNLS-03D3bb) are not discussed. Similarly, the SDSS-II
explicitly removed their peculiar \SNeIa, such as the interacting
SN~2005gj, and the selection cut on the MLCS2k2 light-curve fits
biases their sample against high luminosity, SN~1999aa-like events
\citep{dilday2010,kessler2009}. Because of this, the SNLS and SDSS-II
rates only reflect the frequency for a fraction of the larger
\SNeIa\ population. As to the high luminosity \SNeIa\ excluded by SNLS
and SDSS-II, however, we note that if these constituted more than a
few percent of the total population then they would completely
dominate our flux limited sample, which is not the case.

Although the significance of our rates being higher than the
canonical value is low, it is worth discussing how this may have come
about. One point of intrigue is our excess of \SNeIa\ in dwarf
galaxies over what would be expected from the LOSS or SDSS-II
studies. If our rate for these hosts is correct, then either the
\SNeIa\ rate size relation increases for the lowest luminosity hosts
as compared to the range studied by LOSS, or the faint end of the
galaxy luminosity function used to convert the LOSS luminosity
specific rate to a volumetric rate may be under-counting the actual
supply of dwarf galaxies. 

Such correction factors are not required for the SDSS-II rate, so the
higher frequency of \SNeIa\ in low luminosity hosts is puzzling. In
the full background sample, 7 out of 23 of the ROTSE-IIIb hosts are
fainter than $M_B \ge -16$\,mag, as are 5 out of the 18 used for the
rate calculation with the augmented LOSS pAMD. These fractions
($\sim30$\%) are significantly larger than the 1 out of 79
\SNeIa\ hosts in the SDSS-II $z<0.15$ sample that have such low
luminosity hosts (hosts brighter than $M_r\sim-15$\,mag can be
detected through our analysis out to this redshift limit, but only
SDSS-II SN13038 is fainter). If this difference is attributed to a
selection bias against giant hosts in the ROTSE-IIIb sample, then we
must increase our rate to account for the missing \SNeIa\ in
giants. But to achieve the same low to high-luminosity host ratio as
the SDSS-II, our already high rates must be increased by factors of 4
to 5! 

Another possibility to consider is that the SDSS-II may somehow be
biased against low luminosity hosts, but this is difficult to
imagine. A remote possibility is that the use of host photo-z's in
selecting targets for spectroscopic follow-up may boost the likelihood
of classifying \SNeIa\ in giants rather than in dwarfs, which were
unlikely to have been detected in the reference data available during
the search. In that case, there would still need to be a second bias
to reject the \SNeIa\ in dwarfs since photometrically probable yet
non-spectroscopically confirmed \SNeIa\ were included in the SDSS-II
rates \citep{dilday2010}.

A possible contributor to the relative infrequency of dwarf hosts in
the SDSS-II sample may be the cut on the MLCS2k2 light-curve fits used
in selecting photometrically probably \SNeIa\ \citep{dilday2010}. This
cut can remove high luminosity \SNeIa, such as SN~1999aa, which tend
to be associated with lower luminosity hosts
\citep{kessler2009,sullivan2006}. However, only one of the five
\SNeIa\ from our rate sample that fall in $M_B \ge -16$\,mag hosts can
be considered highly luminous: SN~2007if. Since the SDSS-II does
contain a number of higher redshift \SNeIa\ whose hosts are not
detected even in our deep coadds, it does not appear that the SDSS-II
had a selection bias against either ``hostless'' \SNeIa\ or \SNeIa\ in
dwarf hosts, and it is not obvious why such hosts are so much more
frequent in the ROTSE-IIIb sample.

Another questions is why is there such a larger fraction of faint
(pseudo absolute magnitude $M_R > -18.8$) \SNeIa\ in the LOSS sample
as compared to the SDSS-II?  Despite being targeted, the LOSS galaxy
sample spans a range of masses and colors, and the full LOSS galaxy
sample has absolute B and K-band magnitude distributions that are
consistent with the SDSS-II \SNeIa\ hosts in the $z<0.09$ range (see
Fig.~\ref{fig:hosts}). Applying the LOSS rate-size relation only to
galaxies brighter than $M_K<-20$ in the \citet{kochanek2001} galaxy
luminosity function and selecting only the 48\% of \SNeIa\ fainter
than $M_R>-18.8$ in the LOSS pseudo-absolute magnitude distribution,
the {\it minimum} rate of low-luminosity \SNeIa\ implied is
$0.4\times10^{-5}$\,\SNeIa\,Mpc$^{-3}$\,yr$^{-1}$\,$h_{71}^{3}$. In
this case, the first year SDSS-II $z<0.12$ sample should contain at
least 6 \SNeIa\ fainter than $-18.8$\,mag. Even with the peculiar
SN\,2005hk and the photometrically selected SDSS-II SN09266 (with its
heavy host absorption of $A_V \sim 4$\,mag), there are still only 3
events in the SDSS-II $z<0.12$ first year sample at such low
luminosities.

It is not clear how the SDSS-II could have missed the faint tail of
the \SNeIa\ distribution (e.g. SN~1991bg-like events around $M_B \sim
-17.2$), yet we know of only two such spectroscopically confirmed
events discovered in the three years of the SDSS-II (2007jh and
2007mm; \citealt{mosher2012}) and only one possible SN\,2002cx-like
event (SN\,2007ie; \citealt{ostman2011}). Additionally, the Palomar
Transient Factory \citep{rau2009,law2009} has recently shown that
there are perhaps even more varieties of peculiar, low-luminosity
objects that may be spectroscopically classified as \SNeIa\ 
\citep[e.g.][]{sullivan2011,maguire2011}, but these are apparently
absent from the SDSS-II as well. The apparent lack of low-luminosity
\SNeIa\ could mean that there was an unknown selection bias against
such objects and that the total rate of \SNeIa\ is higher than found
by the SDSS-II.

The tension between the LOSS and SDSS-II pAMDs poses a potentially
worrisome systematic bias for \SNeIa\ rate studies. \citet{dilday2008}
argue that the SDSS-II was sensitive to even sub-luminous
\SNeIa\ below $z<0.12$, so even if their model distribution were
flawed, this would not have a significant impact on their rate
calculation (but again, the dearth of low-luminosity \SNeIa\ as
compared to the expectations of LOSS is puzzling at best and may
suggest the SDSS-II has altogether missed some \SNeIa). However, if
there is a low-luminosity population of \SNeIa\ that is not accounted
for in the SDSS-II model distribution, this will lead to an
overestimation of the search efficiency at the higher redshifts
considered in \citet{dilday2010}, and thus an underestimation of the
actual rate. In general, rate studies that push their samples out to
the highest redshifts possible with their surveys have an even
stronger Malmquist bias and are thus more sensitive to systematic
errors in their adopted pseudo-absolute magnitude distributions. These
distributions may also vary with redshift. It is therefore important
to securely determine the true \SNeIa\ pseudo-absolute magnitude
distribution from complete, volume limited surveys as could
potentially be done by the Palomar Transient Factory at low redshifts
and at moderate redshifts with the Hyper-SuprimeCam on Subaru.

Another issue that may lead to bogus rate evolution measurements is
the changing definition of the \SNeIa\ sub population measured by
different surveys. Low to moderate redshift surveys often use light
curve fits to select
\SNeIa\ \citep[e.g.][]{dilday2010,neill2006,perrett2012}, but the
highest redshift studies often have far fewer epochs available due to
the increased follow-up cost, and must therefore rely more heavily on
color selection techniques \citep[][]{graur2011}. It is not clear
that these different cuts capture identical sub populations so that
differences in the rates between different surveys may be taken
directly as evidence for evolution in the \SNeIa\ rate. This question
may be resolved by performing a rolling search similar to SDSS-II but
sensitive to very high redshift \SNeIa, as will be possible with the
Hyper-SuprimeCam Survey.

\acknowledgments ROTSE-III has been supported by NASA grant
NNX-08AV63G and NSF grant PHY-0801007.  The research of JCW is
supported in part by NSF grant AST1109801. This material is based upon
work supported in part by the National Science Foundation under Grant
No. 1066293 and the hospitality of the Aspen Center for Physics.  This
research made use of Montage, funded by the National Aeronautics and
Space Administration's Earth Science Technology Office, Computation
Technologies Project, under Cooperative Agreement Number NCC5-626
between NASA and the California Institute of Technology. Montage is
maintained by the NASA/IPAC Infrared Science Archive.

\bibliographystyle{aa}

\end{document}